\documentclass[12pt]{article}

\usepackage[a4paper, total={18cm, 23cm}]{geometry}

\usepackage{siunitx}
\usepackage{amsmath}
\usepackage{amsfonts}
\usepackage{amssymb}
\usepackage{graphicx}
\usepackage{multirow}
\usepackage{tikz}
\usetikzlibrary{arrows}
\usetikzlibrary{decorations.markings}
\usepackage{soul}
\usepackage{color}
\usepackage{colortbl}
\definecolor{darkgreen}{rgb}{0.05, 0.5, 0.05}
\usepackage{rotating}

\usepackage{natbib}

\begin{document}

\title{\textbf{\Large{Statistical Methods for cis-Mendelian Randomization with Two-sample Summary-level Data}}}
\author{Apostolos Gkatzionis$^{1, 2}$\footnote{Corresponding author. Email: apostolos.gkatzionis@bristol.ac.uk}, Stephen Burgess$^{1, 3}$, Paul J Newcombe$^1$}
\date{\begin{small}$^1$MRC Biostatistics Unit, University of Cambridge, UK. \\
$^2$MRC Integrative Epidemiology Unit, University of Bristol, UK. \\
$^3$Department of Public Health and Primary Care, School of Clinical Medicine, \\ University of Cambridge, UK.\end{small}}
\maketitle

\abstract{Mendelian randomization is the use of genetic variants to assess the existence of a causal relationship between a risk factor and an outcome of interest. Here, we focus on two-sample summary-data Mendelian randomization analyses with many correlated variants from a single gene region, and particularly on cis-Mendelian randomization studies which use protein expression as a risk factor. Such studies must rely on a small, curated set of variants from the studied region; using all variants in the region requires inverting an ill-conditioned genetic correlation matrix and results in numerically unstable causal effect estimates. We review methods for variable selection and estimation in cis-Mendelian randomization with summary-level data, ranging from stepwise pruning and conditional analysis to principal components analysis, factor analysis and Bayesian variable selection. In a simulation study, we show that the various methods have a comparable performance in analyses with large sample sizes and strong genetic instruments. However, when weak instrument bias is suspected, factor analysis and Bayesian variable selection produce more reliable inferences than simple pruning approaches, which are often used in practice. We conclude by examining two case studies, assessing the effects of LDL-cholesterol and serum testosterone on coronary heart disease risk using variants in the \textit{HMGCR} and \textit{SHBG} gene regions respectively.}

\vspace{0.5cm}

\textbf{Keywords:} Mendelian randomization, correlated instruments, pruning, principal components analysis, factor analysis, Bayesian variable selection.

\vspace{0.5cm}

\section{Introduction}

Mendelian randomization (MR) is the use of genetic data in order to assess the existence of a causal relationship between a modifiable risk factor and an outcome of interest \citep{DaveySmith2003, Burgess2015book}. It is an application of instrumental variables analysis in the field of genetic epidemiology, where genetic variants are used as instruments. The approach has received much attention in recent years and has been used to identify a large number of cause-effect relationships in the epidemiologic literature \citep{Boef2015}. For example, MR studies have strengthened the evidence for a causal link between lipoprotein(a) and coronary heart disease (CHD) \citep{Kamstrup2009}, but have weakened the evidence for an effect of C-reactive protein on CHD risk \citep{Ccgc2011}.

A Mendelian randomization analysis requires a set of genetic variants that satisfy the instrumental variable assumptions: genetic variants should be (i) associated with the risk factor of interest, (ii) independent of confounders of the risk factor-outcome association, and (iii) they should only influence the outcome through their effect on the risk factor (the no-pleiotropy assumption). Under these three assumptions, MR offers a framework for assessing whether the risk factor is causally related to the outcome; under additional modelling assumptions, one can also estimate the magnitude of the risk factor-outcome causal effect.

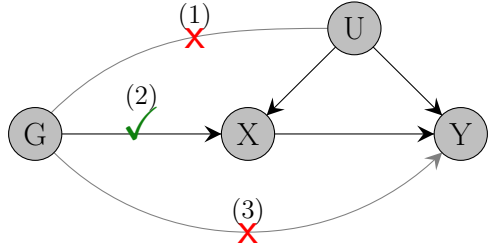
\begin{figure}[bt]
\centering
\begin{tikzpicture}[scale=0.7, every node/.style={scale=0.7}]
\draw [decoration={markings,mark=at position 1 with
    {\arrow[scale=2,>=stealth]{>}}},postaction={decorate}] (-3.5, 0) -- (-0.5, 0);
\draw [decoration={markings,mark=at position 1 with
    {\arrow[scale=2,>=stealth]{>}}},postaction={decorate}] (1.65, 1.65) -- (0.35, 0.35);
\draw [decoration={markings,mark=at position 1 with
    {\arrow[scale=2,>=stealth]{>}}},postaction={decorate}] (2.35, 1.65) -- (3.65, 0.35);
\draw [decoration={markings,mark=at position 1 with
    {\arrow[scale=2,>=stealth]{>}}},postaction={decorate}] (0.5, 0) -- (3.5, 0);
\draw [gray, decoration={markings,mark=at position 1 with
    {\arrow[scale=2,>=stealth]{>}}},postaction={decorate}] (-3.65, -0.35) to [out = 315, in = 225] (3.65, -0.35);
\draw [gray] (1.5, 2) to [out = 180, in = 45] (-3.70, 0.40);
\draw[fill = lightgray] (0, 0) circle (0.5cm);
\draw[fill = lightgray] (-4, 0) circle (0.5cm);
\draw[fill = lightgray] (4, 0) circle (0.5cm);
\draw[fill = lightgray] (2, 2) circle (0.5cm);
\node (a) at (0, 0) {\Large{X}};
\node (b) at (-4, 0) {\Large{G}};
\node (c) at (4, 0) {\Large{Y}};
\node (d) at (2, 2) {\Large{U}};
\node (f) at (-2, 0.2) {\textcolor{darkgreen}{\huge{\checkmark}}};
\node (g) at (-1, 1.82) {\textcolor{red}{\huge{\textsf{x}}}};
\node (h) at (0, -1.87) {\textcolor{red}{\huge{\textsf{x}}}};
\node (j) at (-1, 2.22) {\large{(1)}};
\node (i) at (-2, 0.7) {\large{(2)}};
\node (k) at (0, -1.47) {\large{(3)}};
\end{tikzpicture}
\caption{A causal diagram representation of the three assumptions of Mendelian randomization. Here, $X$ represents the risk factor, $Y$ the outcome, $G$ the genetic instrument and $U$ denotes confounders of the $X-Y$ relationship.} 
\label{dag}
\end{figure}

Traditionally, Mendelian randomization studies are implemented using independent genetic variants from across the whole genome. This is especially the case when studying complex polygenic traits such as body mass index, cholesterol or blood pressure. Such genome-wide analyses rely on published results from large-scale GWAS studies to identify large numbers of genetic regions associated with the risk factor studied. Genetic variants in each region are pruned for independence and only the variant with the smallest p-value, or a small number of weakly correlated variants with small p-values are selected. Variants from different regions are then combined to create a genome-wide set of instruments for MR, thus increasing the power of the analysis to detect a causal relationship. 

In parallel to the traditional genome-wide MR studies described above, recent years have seen an increased number of MR studies based on a single gene region \citep{Schmidt2019, Gill2020}. These single-region investigations are sometimes called cis-MR studies, because they use cis-variants (variants in close proximity to the gene of interest) as instrumental variables. The rise in popularity of cis-MR studies has been driven by the increasing appreciation of the fact that such studies can be used to inform drug target discovery and validation \citep{Gill2020}. Cis-MR studies typically use protein expression (or downstream biomarkers strongly associated with protein expression) as a risk factor, since each coding gene region encodes one protein and since proteins are the drug targets of most medicines. By studying the associations of variants in the region with a disease outcome of interest, cis-MR studies can thus provide information on whether the encoded protein can be used as a drug target for the outcome. Their scope is therefore different to that of genome-wide MR analyses, which aim to assess causal relationships between phenotypic traits, and more similar to that of transcriptome-wide and proteome-wide association studies, as we discuss later.

Cis-MR studies typically rely on a single gene region, and therefore have to select genetic instruments among a potentially large number of correlated variants. Using all variants in the region can result in numerically unstable causal effect estimates \citep{Burgess2017}, so instead a small set of variants that are highly informative for the risk factor is normally used. Through the existing cis-MR literature, a variety of techniques have been employed to select these variants. However, to date no review or comparison of the different methods has been published, and applied cis-MR analyses typically use some form of LD-pruning for variable selection. The aim of this paper is to provide an overview of statistical methods for cis-MR studies, focusing on the common case of two-sample MR studies with summary-level data. After reviewing commonly used approaches such as LD-pruning and conditional analysis, we discuss the use of principal components analysis (PCA), factor analysis and stochastic-search variable selection. The use of PCA \citep{Burgess2017} was proposed for MR with correlated instruments and can be directly extended to cis-MR. The use of factor analysis was recently explored by \cite{Patel2020}, building on similar methods for instrumental variables analysis with individual-level data \citep{Bai2002}. Stochastic search can be implemented using the JAM algorithm (joint analysis of marginal summary statistics), originally proposed by \cite{Newcombe2016} for fine-mapping densely genotyped regions and recently adapted for MR \citep{Gkatzionis2019}. We discuss how to implement each method in the context of cis-MR and compare their performance in simulation studies.

We also provide two real-data applications, investigating associations of variants in the \textit{SHBG} and \textit{HMGCR} regions with coronary heart disease risk. The \textit{HMGCR} gene encodes the protein HMG-coenzyme A reductase, which is known to play an important role in the cholesterol-biosynthesis pathway. Its inhibition with statins is known to reduce LDL-cholesterol levels, therefore statin treatment is often prescribed to individuals in high CHD risk. The \textit{SHBG} region encodes the protein sex-hormone binding globulin, and is known to contain genetic associations with testosterone and other sex hormones; there is some uncertainty on whether testosterone levels are associated with CHD risk, although a recent MR study has suggested no causal relationship \citep{Schooling2018}.

\section{An Overview of Cis-Mendelian Randomization}

Drug development is a challenging and costly process with uncertain results, and the failure rate for drugs in clinical development can be as high as $90\%$ \citep{Hay2014}. Since proteins are typically the targets of common drugs, cis-Mendelian randomization uses expression data for druggable proteins as risk factors, in order to aid in drug target discovery and validation. The wealth of available genetic data, combined with the ability to routinely implement MR analyses using existing software \citep{Walker2019mrbase, Yavorska2017} makes cis-MR a useful approach to complement and inform drug development. This can be evidenced by the increased success rate of drug development programs with genomic support \citep{Nelson2015}.

Cis-MR studies can be used in order to assess the suitability of proteins as potential drug targets, predict and inform the results of clinical trials, or investigate repurposing opportunities and potential side-effects of existing drugs. For example, a Mendelian randomization analysis of LDL cholesterol and cardiovascular disease risk based on the \textit{ACLY} region \citep{Ference2019} was recently used alongside a clinical trial \citep{Ray2019} to assess the suitability of the \textit{ACLY} region as a potential drug target for lowering LDL-cholesterol. Similar analyses have been conducted using the \textit{CETP} region to instrument lipid fractions and blood pressure \citep{Sofat2010}, using the \textit{IL6R} region to assess the effect of interleukin 6 reduction on coronary heart disease risk \citep{Swerdlow2012}, and using the \textit{HMGCR} \citep{Swerdlow2015} and \textit{PCSK9} regions \citep{Schmidt2017} to explore potential effects of LDL-cholesterol lowering treatments on body mass index and type 2 diabetes risk.

Ideally, cis-MR should be conducted using protein expression data for the target protein risk factor \citep{Gill2020}. However, data on protein expression are not always available, and researchers often use either gene expression data for the associated gene region or genetic associations with a downstream biomarker instead. For example, LDL-cholesterol has been used as a downstream risk factor in order to study the effects of HMGCR inhibition on the risk of type 2 diabetes \citep{Swerdlow2015}.

Genetic variants for cis-MR are selected from a region containing the protein-encoding gene. The gene region is defined to contain the target gene, as well as variants within a few hundred thousand base pairs on either side of the gene, in order to include possible cis-eQTL variants. In principle, MR of protein expression can be implemented using variants from across the genome. However, since cis-MR is used for drug target discovery and validation, it is more relevant to focus on the protein-encoding gene region \citep{Schmidt2019}. This is especially the case when using a downstream biomarker as a proxy for protein expression, as there may be variants associated with the biomarker but not with the target protein. In the LDL-cholesterol example mentioned above, using variants associated with LDL-cholesterol across the genome would implicate multiple causal mechanisms affecting LDL-cholesterol concentration, many of which may not be related to the function of the HMGCR protein. Therefore, if the aim of the analysis is to assess potential side-effects of HMGCR-inhibiting drugs, it would be better to conduct the analysis using only variants in the \textit{HMGCR} region. In addition, variants in a single gene region often exhibit less heterogeneity in their genetic effects than variants from across the genome \citep[see Figure 4 of][for example]{Burgess2017sensitivity}.

Within the limits of a gene region, it is often possible to identify hundreds of correlated variants marginally associated with protein expression. Researchers must then select which of these variants to incorporate in a subsequent MR analysis; including all available variants is usually avoided as it can introduce numerical approximation errors in estimating the MR causal effect \citep{Burgess2017}. Some authors have suggested prioritizing variants based on functional annotations; however, \cite{Schmidt2019} investigated this strategy in a simulation study and found little to no difference in causal inferences conducted using functional variants compared to non-coding variants. Therefore, the selection of genetic variants is usually performed based on the strength of their associations with the risk factor, while accounting for LD correlation to reduce numerical approximation errors. In the next section, we review statistical approaches for implementing this variable selection procedure.

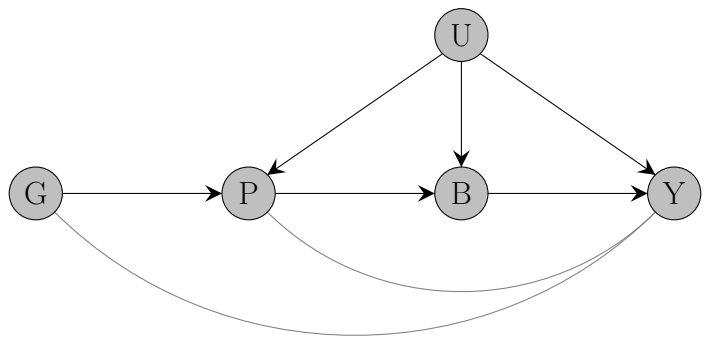
\begin{figure}[bt]
\centering
\begin{tikzpicture}[scale=0.7, every node/.style={scale=0.7}]
\draw [decoration={markings,mark=at position 1 with
    {\arrow[scale=2,>=stealth]{>}}},postaction={decorate}] (-5.5, 0) -- (-2.5, 0);
\draw [decoration={markings,mark=at position 1 with
    {\arrow[scale=2,>=stealth]{>}}},postaction={decorate}] (-1.5, 0) -- (1.5, 0);
\draw [decoration={markings,mark=at position 1 with
    {\arrow[scale=2,>=stealth]{>}}},postaction={decorate}] (2.5, 0) -- (5.5, 0);
\draw [decoration={markings,mark=at position 1 with
    {\arrow[scale=2,>=stealth]{>}}},postaction={decorate}] (2, 2.5) -- (2, 0.5);
\draw [decoration={markings,mark=at position 1 with
    {\arrow[scale=2,>=stealth]{>}}},postaction={decorate}] (1.65, 2.65) -- (-1.65, 0.35);
\draw [decoration={markings,mark=at position 1 with
    {\arrow[scale=2,>=stealth]{>}}},postaction={decorate}] (2.35, 2.65) -- (5.65, 0.35);
\draw [gray, decoration={markings,mark=at position 1 with
    {\arrow[scale=2,>=stealth]{>}}},postaction={decorate}] (-5.65, -0.35) to [out = 315, in = 225] (5.65, -0.35);
\draw [gray, decoration={markings,mark=at position 1 with
   {\arrow[scale=2,>=stealth]{>}}},postaction={decorate}] (-1.65, -0.35) to [out = 315, in = 225] (5.65, -0.35);
\draw[fill = lightgray] (-6, 0) circle (0.5cm);
\draw[fill = lightgray] (-2, 0) circle (0.5cm);
\draw[fill = lightgray] (2, 0) circle (0.5cm);
\draw[fill = lightgray] (6, 0) circle (0.5cm);
\draw[fill = lightgray] (2, 3) circle (0.5cm);
\node (a) at (-6, 0) {\Large{G}};
\node (b) at (-2, 0) {\Large{P}};
\node (c) at (2, 0) {\Large{B}};
\node (c) at (6, 0) {\Large{Y}};
\node (d) at (2, 3) {\Large{U}};
\end{tikzpicture}
\caption{Directed acyclic graph illustrating why pleiotropic bias is less likely in MR studies of protein expression than in studies using downstream phenotypic traits as risk factors. Here, $G$ denotes the genetic instrument, $P$ protein expression, $B$ the downstream biomarker, $Y$ the outcome and $U$ denotes confounding factors. Both types of MR studies are affected by pleiotropy due to ``direct" effects of the genetic instrument on the outcome ($G \rightarrow Y$ pathway), but standard MR analyses are also subject to pleiotropy due to potential effects of the protein on the outcome ($G \rightarrow P \rightarrow Y$ pathway) which is not the case for studies of disease progression.}
\label{pleiotropy}
\end{figure}

Finally, similar to standard Mendelian randomization analyses, genetic variants used for cis-MR must satisfy the instrumental variables assumptions. Violations of the instrumental variable assumptions are often a concern because they have the potential to devalidate tests of causal association between the risk factor and the outcome and to bias causal effect estimates. In principle, violations of the no-pleiotropy assumption are less likely to occur when studying protein expression, because proteins are causally upstream of metabolites and biomarkers which constitute the most common risk factors used in traditional MR analyses \citep{Swerdlow2016, Schmidt2019}. This is illustrated in the causal diagram of Figure~\ref{pleiotropy} \citep[which is adapted from Figure 1 of][]{Schmidt2019}. 

Nevertheless, pleiotropic bias remains a concern even for studies of protein expression; it would arise if variants from the same region were involved in different biological pathways relevant to the outcome, resulting in heterogeneous causal effect estimates. When pleiotropy does occur, standard algorithms for pleiotropy-robust MR (including the JAM algorithm) can be used to address it. However, care must be taken when extrapolating the assumptions made by such algorithms to the framework of many correlated genetic variants. For example, suppose that a region contains a single pleiotropic variant, but that variant is strongly correlated with half of the other variants in the region. This will induce bias in the marginal causal effect estimates of the correlated variants, and hence pleiotropy-robust methods requiring that a majority or plurality of genetic variants in the region are valid instruments will be biased.

\section{Statistical Methods for cis-Mendelian Randomization}

\subsection{Notation}

We now present various approaches for selecting genetic variants and computing causal effect estimates in cis-MR studies. Let $X$ denote the risk factor of interest, $Y$ denote the outcome, $G_1, \dots, G_P$ denote a set of $P$ putative genetic instruments from a single gene region, and let $U$ represent confounders of the risk factor-outcome association. Also, let $\theta$ denote the causal effect of the risk factor on the outcome. Our task is to obtain a subset of the genetic variants to use as instruments, an estimator $\hat{\theta}$ of the causal effect parameter and an associated standard error $\text{se} (\hat{\theta})$.

We work under the framework of two-sample summary-data MR and assume the availability of two sets of summary statistics, one for SNP-risk factor and one for SNP-outcome associations, obtained from different datasets. We use $\hat{\beta}_{Xj}$, $\hat{\beta}_{Yj}$ to denote the marginal associations between genetic variant $G_j$ and the two traits, and $\hat{\sigma}_{Xj}^2$, $\hat{\sigma}_{Yj}^2$ for the corresponding standard errors. In addition, $p_{Xj}$ denotes the p-value of association between variant $G_j$ and the risk factor. Finally, we assume that we have access to a reference dataset, such as $1000$ genomes or the UK Biobank, from which to estimate correlations between genetic variants.

\subsection{Single-Variant MR}

Some studies in the cis-MR literature have used only the variant with the smallest univariate p-value in a region as an instrument \citep{Sofat2010, Swerdlow2015}. With a single genetic variant, the MR causal effect can be estimated using a ratio formula. This avoids the need to account for genetic correlations and to perform complex numerical tasks such as inverting the genetic covariance matrix. On the other hand, if the region contains multiple variants with independent associations with the risk factor, a single-variant MR analysis will not be able to model all the genetic effects in the region. This typically translates into a loss of power for the cis-MR analysis. In addition, identification of the smallest p-value variant in the region can be subject to winner's curse bias. Therefore, most recent cis-MR papers have utilised variable selection or dimension reduction approaches in order to include several variants per region.

\subsection{LD-Pruning}

Perhaps the most common approach in the literature for selecting genetic variants for inclusion into a cis-MR study is LD-pruning \citep{Dudbridge2015, Schmidt2019}. Given a set of candidate genetic variants $G_1, \dots, G_P$, the stepwise pruning procedure iterates between:
\begin{itemize}
	\item Selecting the variant with the smallest p-value and including it in the analysis.
	\item Identifying all variants whose correlation with the previously selected variant exceeds a correlation threshold $\rho$ and removing them from further consideration.
\end{itemize}
This is repeated until all the candidate variants have either been selected for inclusion or discarded. Alternatively, the process may be stopped when there are no remaining variants with p-values lower than a significance threshold $\tau$, in order to guard against the inclusion of weak instruments into the analysis. The threshold $\tau$ is often taken to be the GWAS significance threshold $\tau = 5 \times 10^{-8}$, although there is some evidence that using more relaxed thresholds may be beneficial \citep{Dudbridge2013}. 

Once LD-pruning is implemented and a suitable set of genetic instruments is selected, the MR causal effect can be estimated using an inverse variance weighted (IVW) estimate. If the correlation threshold $\rho$ is small enough, the selected variants can be considered approximately independent and the standard IVW estimate, 
\begin{eqnarray}
	\hat{\theta}_{ivw} & = & \frac{\sum_{j = 1}^P \hat{\beta}_{Xj} \hat{\beta}_{Yj} \hat{\sigma}_{Yj}^{-2}}{\sum_{j = 1}^P \hat{\beta}_{Xj}^2 \hat{\sigma}_{Yj}^{-2}} \nonumber \\
 	\text{se} (\hat{\theta}_{ivw})^2 & = & \frac{1}{\sum_{j = 1}^P \hat{\beta}_{Xj}^2 \hat{\sigma}_{Yj}^{-2}} \nonumber
\end{eqnarray}
can be used. This approach is typically used in genome-wide MR analyses and can also be adapted to account for uncertainty in G-X associations \citep{Bowden2018}. However, for single-region MR, the use of a small correlation threshold may result in the exclusion of causal variants, especially when studying gene regions with multiple independent associations with the risk factor. On the other hand, if a larger threshold is used, the standard IVW estimate will underestimate the causal standard error, since it does not adjust for correlations. In this case it is better to use a modified version of of the IVW estimator that takes genetic correlations into account \citep{Burgess2016}:
\begin{eqnarray}
	\hat{\theta}_{cor} & = & \left( \hat{\beta}_X^T \Omega^{-1} \hat{\beta}_X \right)^{-1} \left( \hat{\beta}_X^T \Omega^{-1} \hat{\beta}_Y \right) \label{ivw.cor} \\
	\text{se} (\hat{\theta}_{cor})^2 & = & \left( \hat{\beta}_X^T \Omega^{-1} \hat{\beta}_X \right)^{-1} \label{ivw.cor.se}
\end{eqnarray}
where $\Omega$ denotes the genetic covariance matrix, $\Omega_{i, j} = \text{Cov} (G_i, G_j)$, and is usually estimated from a reference dataset. The estimator \eqref{ivw.cor} can be motivated from the literature on meta-analysis. It is sometimes referred to as the generalized least squares method \citep{Schmidt2019} because it can be derived using generalized linear regression to model the genetic associations with the outcome in terms of the genetic associations with the risk factor \citep{Burgess2016, Burgess2017}.

The correlated-instruments IVW estimator requires the genetic covariance matrix $\Omega$ to be computed and inverted, a process which can become numerically unstable if highly correlated variants are used \citep{Burgess2017}. This is the main reason why cis-MR analyses cannot use all available variants within a region. Numerical instabilities are less likely to occur when the IVW estimator is used in combination with pruning, since the pairwise correlations of variants selected by LD-pruning are all lower than the specified correlation threshold $\rho$, although problems can still occur if the genetic covariance matrix is near-singular.

A potential criticism for the LD-pruning approach is that there is no consensus in the literature on how to select the correlation threshold. Implementations of the method using different correlation thresholds can give rise to different results, especially when large thresholds are used \citep{Burgess2017}. To address this issue, \cite{Schmidt2019} proposed that stepwise pruning should be implemented with a variety of pruning thresholds, as a form of sensitivity analysis. In addition, stepwise methods for variable selection have been criticized for being unstable and getting stuck in local maxima of the model space \citep{Hocking1976, Tibshirani1996, Miller2002} and more flexible variable selection approaches are often preferred in some applications of biostatistics \citep{Newcombe2016, Benner2016, Asimit2019}, although it is not clear whether the added flexibility provides a substantial benefit in Mendelian randomization studies.

\subsection{Conditional Analysis}

Conditional analysis is a process similar to LD-pruning, except variant selection is based on genetic associations with the risk factor conditional on any previously selected variants instead of univariate marginal p-values. Perhaps the most established algorithm for conditional analysis is the Conditional and Joint (CoJo) algorithm of \cite{Yang2012}. The algorithm aims to identify a set of selected variants, to be included in the MR analysis. Initially, this set only contains the genetic variant with the smallest univariate p-value in the region. The algorithm then executes the following steps in each iteration:
\begin{itemize}
	\item Calculate the p-values of all candidate genetic variants conditional on the selected variants. We refer to \cite{Yang2012} for more details on how to do this. As with stepwise pruning, the p-values only need to be computed for genetic variants whose correlations with previously selected variants do not exceed a pre-specified correlation threshold $\rho$.
	\item Identify the variant with the smallest conditional p-value (provided that this p-value is smaller than a significance threshold $\tau$) and add it to the set of selected variants.
	\item Regress the risk factor on all selected variants and compute the joint-model p-values for each selected variant.
	\item If any of the joint-model p-values is larger than the significance threshold $\tau$, drop the variant with the largest p-value from the set of selected variants.
\end{itemize}
The algorithm is run until it can no longer add or remove any variants. Using conditional (and not marginal) p-values for variable selection can improve the algorithm's ability to identify variants with independent signals compared to LD-pruning. Upon finishing, the algorithm provides a set of genetic variants which can subsequently be used to conduct Mendelian randomization. A causal effect estimate based on the selected variants is then obtained using correlated-instruments IVW \eqref{ivw.cor}-\eqref{ivw.cor.se}.

The conditional and joint-model p-values can be computed using marginal summary statistics \citep{Yang2012}. The fact that the algorithm only requires summary-level data, along with its implementation as part of the GCTA software, have contributed to conditional analysis becoming a popular option for selecting variants from single genetic regions.

In practice, conditional analysis often has a similar performance to LD-pruning \citep{Dudbridge2015}. Both approaches rely on a correlation threshold parameter $\rho$ and can produce inconsistent results for different values of the correlation parameter, especially for large values (e.g. $\rho \geq 0.9$), due to numerical instabilities.

\subsection{Principal Component Analysis}

Principal components analysis (PCA) is widely used in GWAS studies to adjust for population stratification. In the context of correlated-instruments summary-data MR, the use of principal components was proposed by \cite{Burgess2017}. The method aims to identify linear combinations of genetic variants that are orthogonal to each other and explain as much of the variance in the genetic dataset as possible. These linear combinations are then used as genetic instruments to assess the causal relationship between the risk factor and outcome, using an IVW estimate.

Principal components analysis is applied to a matrix $\Psi$ with elements
\begin{equation*}
	\Psi_{ij} = \hat{\beta}_{Xi} \hat{\beta}_{Xj} \hat{\sigma}_{Yi}^{-1} \hat{\sigma}_{Yj}^{-1} \rho_{ij} = \hat{\beta}_{Xi} \hat{\beta}_{Xj} \hat{\sigma}_{Yi}^{-2} \hat{\sigma}_{Yj}^{-2} \Omega_{ij}
\end{equation*}
where $\rho_{ij} = \text{Cor} (G_i, G_j)$. The matrix $\Psi$ is practically a transformed version of the genetic correlation matrix $R = (\rho_{ij})$. It is preferred over the untransformed matrix because its entries depend on the precisions $\hat{\sigma}_{Yj}^{-1}$ of univariate estimates and thus, if two genetic variants are almost perfectly correlated, the variant with the highest precision will be prioritized.

Principal components analysis uses an eigendecomposition for the matrix $\Psi$ and identifies the diagonal matrix of eigenvalues $\Lambda = (\lambda_{ii})$ and the corresponding matrix of eigenvectors $W$ so that $\Psi = W^T \Lambda W$. By transforming the summary statistics
\begin{equation*}
	\bar{\beta}_X = W^T \hat{\beta}_X \;\;\; , \;\;\; \bar{\beta}_Y = W^T \hat{\beta}_Y
\end{equation*}
and the genetic covariance matrix
\begin{equation*}
	\bar{\Omega} = W^T \Omega W
\end{equation*}
one can rewrite the IVW estimate \eqref{ivw.cor} in terms of these transformed vectors:
\begin{equation}
	\hat{\theta}_{cor} = \left( \bar{\beta}_X^T \bar{\Omega}^{-1} \bar{\beta}_X \right)^{-1} \left( \bar{\beta}_X^T \bar{\Omega}^{-1} \bar{\beta}_Y \right) \label{pca}
\end{equation}
Evaluating \eqref{pca} is subject to the same numerical difficulties as \eqref{ivw.cor}. However, one can construct a numerically stable approximation for $\hat{\theta}_{cor}$ by using only the first $k$ principal components. If $\tilde{W}$ is the sub-matrix of $W$ containing only the first $k$ columns and  $\tilde{\beta}_{X} = \tilde{W}^T \hat{\beta}_X$, $\tilde{\beta}_{Y} = \tilde{W}^T \hat{\beta}_Y$, $\tilde{\Omega} = \tilde{W}^T \Omega \tilde{W}$, then the causal effect estimate based on the first $k$ principal components is
\begin{equation*}
	\hat{\theta}_{cor} \approx \left( \tilde{\beta}_{X}^T \tilde{\Omega}^{-1} \tilde{\beta}_{X} \right)^{-1} \left( \tilde{\beta}_{X}^T \tilde{\Omega}^{-1} \tilde{\beta}_{Y} \right)
\end{equation*}
The number $k$ of principal components to use is a tuning parameter for the method. In practice, it is often specified so that the selected principal components explain a specific proportion of variation in the genetic dataset. The criterion used for MR \citep{Burgess2017} is to select principal components that explain either $99\%$ or $99.9\%$ of the variation in the data.

\subsection{The JAM Algorithm}

The JAM algorithm \citep[joint analysis of marginal summary statistics,][]{Newcombe2016} is a Bayesian stochastic-search variable selection algorithm that was proposed for the purpose of statistical fine-mapping from summary GWAS results. The algorithm aims to identify genetic variants robustly and independently associated with a trait of interest, among a large set of correlated candidate variants. JAM only requires the availability of GWAS summary-level data, and genetic correlations which can be estimated from an external reference dataset. Using these summary data, JAM identifies sets of genetic variants that are most plausibly associated with the trait of interest.

The algorithm assumes a linear regression model for the trait $X$ in terms of genetic variants $G = (G_1, \dots, G_P)$:
\begin{equation}
	X \sim N \left(G b, \sigma_X^2 \right) \label{jam.x}
\end{equation}
where $b$ denotes the vector of joint effects of the genetic variants on the trait (in contrast to $\beta_X, \beta_Y$, which represent marginal effects of each variant separately). In order to fit this regression model using summary statistics, JAM employs a transformation $z = G^T X$, for which \eqref{jam.x} implies that
\begin{equation}
	z \sim N \left(G^T G b, G^T G \sigma_X^2 \right) \label{jam.z}
\end{equation}
Assuming Hardy-Weinberg equilibrium and that the trait measurements and genetic data have been centered, each element $z_j$ of the vector $z$ can be approximated using the marginal summary statistics $\hat{\beta}_{Xj}$ and effect allele frequencies $f_j$,
\begin{equation*}
	z_j = 2 f_j (1 - f_j) N_1 \hat{\beta}_{Xj}
\end{equation*}
where $N_1$ is the sample size from which the G-X associations are inferred. In practice, an additional Cholesky decomposition is implemented to avoid modelling under the correlated error structure of \eqref{jam.z}; we refer to \cite{Newcombe2016} for the technical details. In addition, note that $G^T G$ is ($N_1 - 1$ times) the genetic covariance matrix, which can be estimated from reference data. Hence, equation \eqref{jam.z} can be used to construct a summary-data likelihood $\ell (z, G_{ref} | b, \sigma_X^2)$. A similar likelihood is used by the conditional and joint method \citep{Yang2012} to conduct the joint analysis step. JAM employs Bayesian analysis instead, and uses conjugate normal-inverse-gamma g-priors $p(b, \sigma_X^2)$ to facilitate Bayesian inference for the genetic associations with the trait.

The likelihood \eqref{jam.z} relies on a fixed set of genetic variants. For variable selection, we assume that the likelihood has been conditioned on a specific set of variants $\gamma$ and use a Beta-Binomial prior $p(\gamma)$ for the model coefficient in order to obtain the posterior 
\begin{equation}
	p (\gamma, b, \sigma_X^2 | z, G_{ref}) \propto \ell(z, G_{ref} | b, \sigma_X^2, \gamma) p(b, \sigma_X^2 | \gamma) p(\gamma) \label{jam.post}
\end{equation}
This posterior is difficult to evaluate analytically, but JAM implements stochastic-search model selection via a reversible-jump Markov Chain Monte Carlo (RJMCMC) algorithm to sample from \eqref{jam.post} and identify the most suitable sets of genetic variants. The stochastic search procedure starts from a set $\gamma_{(0)}$ containing only the variant with the smallest p-value. Then in each iteration, given a current set of genetic variants $\gamma_{(k)}$, JAM randomly proposes a new set $\gamma_{(k + 1)}$ by either
\begin{itemize}
	\item Adding a new variant to the current set $\gamma_{(k)}$.
	\item Removing a variant from $\gamma_{(k)}$.
	\item Swapping a variant in $\gamma_{(k)}$ for a variant not in $\gamma_{(k)}$.
\end{itemize}
The algorithm then decides whether to accept the new set by evaluating how well it fits the trait, according to the posterior \eqref{jam.post}. The process is repeated for a large number of iterations, and the various sets of variants are assigned posterior probabilities according to how often they were visited. The stochastic search procedure allows JAM to evaluate a wide range of different models and efficiently explore the parameter space of genetic configurations.

If the trait studied is binary, $z$ can be derived by mapping univariate log-odds ratios to the univariate effects that would have been estimated if the binary trait was modelled by linear regression, as has been done in other linear-based summary data algorithms \citep{Vilhjalmsson2015}.

For Mendelian randomization, the above procedure can be used to identify variants strongly and independently associated with the risk factor of interest \citep{Gkatzionis2019}. An IVW causal effect estimate can then be obtained for each set of variants visited by JAM, and these model-specific estimates can be combined into an aggregate causal effect estimate by model averaging.

Since it jointly models all available genetic variants and can account for genetic correlations, JAM is naturally suited for cis-MR. The algorithm discards variants which are associated with the risk factor only through their correlation with other present variants, and its posterior models include variants independently associated with the risk factor. JAM is not completely free of numerical issues when implemented on near-collinear variants, but these issues can be overcome by employing pruning as a first step before calling the algorithm. An advantage over the LD-pruning approach is that JAM is more robust to the choice of pruning thresholds, since it implements a second layer of variable selection after pruning is finished. The pruning step can be implemented using a high correlation threshold (e.g. $\rho = 0.9$) and no significance threshold. As an alternative, JAM can avoid the need for pruning by using a ridge term for the genetic covariance matrix to make its inversion more stable. The choice of a pruning threshold is then replaced by the tuning of the ridge term parameter. Here we focus on implementations using pruning as a first step and do not explore the use of a ridge term.

Outside of the context of cis-MR, JAM has been extended to handle violations of the no-pleiotropy assumption when working with multiple independent genetic variants, by augmenting its variable selection with a heterogeneity loss function to penalize and downweight pleiotropic variants \citep{Gkatzionis2019}. Furthermore, a hierarchical version of the algorithm that is useful for multivariable MR, as well as transcriptome-wide association studies, was recently developed by \cite{Jiang2020}.

\subsection{Factor-based Methods}

Recently, \cite{Patel2020} proposed the use of factor analysis for MR with correlated weak instruments. The authors model the genetic variants in terms of a factor model,
\begin{equation}
	G = \Lambda f + \epsilon_f
\end{equation}
where $f$ is a vector of $k < P$ latent factors, $\Lambda$ is a matrix of factor loadings and $\epsilon_f$ is a vector of idiosyncratic errors. The matrix $\Lambda$ can be estimated using the first $k$ eigenvectors in the eigendecomposition of the (rescaled) genetic covariance matrix $G^T G$. This covariance matrix and its eigendecomposition can be approximated if one has access to a reference dataset. The number $k$ of latent factors can be selected either empirically, as is common in factor analysis, or in a more disciplined way, for example using the factor penalization method of \cite{Bai2002}.

The factors are then used as genetic instruments in a standard MR model:
\begin{eqnarray}
	X & = & b_X^T f + \epsilon_X \nonumber \\
	Y & = & b_Y^T f + \epsilon_Y \nonumber \\
	\beta_Y & = & \theta \beta_X \nonumber
\end{eqnarray}
where $\epsilon_X, \epsilon_Y$ are mean-zero error terms. An estimate of the causal effect parameter $\theta$ can be obtained by limited-information maximum likelihood (LIML), minimizing the criterion
\begin{equation*}
	\hat{g} (\theta) = \frac{1}{P} \hat{\Lambda}^T \hat{D}_Y \hat{\beta}_Y - \theta \frac{1}{P} \hat{\Lambda}^T \hat{D}_X \hat{\beta}_X
\end{equation*}
where $\hat{\Lambda}$ is the estimated matrix of factor loadings and $\hat{D}_X, \hat{D}_Y$ are the diagonal matrices whose diagonal terms are the sample variances of each genetic instrument; these can be approximated from the reference dataset. Minimization of (a weighted version of) $\hat{g} (\theta)$ yields the factor LIML (F-LIML) estimate.

Under certain conditions, the F-LIML estimator has been shown to be consistent and asymptotically normal \citep{Patel2020}. Unfortunately, these conditions are often violated under a weak instruments framework and F-LIML may suffer from weak instrument bias. In this scenario, an alternative is to conduct hypothesis tests, using the factors as genetic instruments, in order to assess the existence of a causal effect. Such tests were first developed in the instrumental variables literature and adapted to work with summary-level data \citep{Patel2020}. They include the Anderson-Rubin \citep{Anderson1949}, Lagrange multiplier \citep{Kleibergen2002} and conditional likelihood ratio \citep{Moreira2003} tests. The various tests are more robust to weak instrument bias than the F-LIML estimator, and were shown to have higher power than the principal components method under a weak instruments setting.

\subsection{Separating Variable Selection from Estimation}

The approaches discussed in this section implement both variable selection and causal effect estimation. Here, we have focused on the variable selection part, because this is where most differences between the various approaches are observed. In terms of variable selection, LD-pruning and conditional analysis implement stepwise selection of genetic variants, JAM implements Bayesian stochastic search, and PCA and factor analysis construct linear combinations using all available variants and use those linear combinations as instruments.

We have coupled each variable selection algorithm with the estimation method most commonly associated with it in practice, or (for newly-developed algorithms) with the estimation method proposed by its authors. For top-SNP analysis, causal effect estimation is performed using a Wald ratio estimate. For pruning, conditional analysis, PCA and JAM, we have used the popular IVW formula. For the factor-based approach, \cite{Patel2020} explicitly proposed using likelihood-based methods (LIML and the conditional likelihood ratio test) for estimation using the factors. These three methods (Wald ratio, IVW and LIML) are the most common estimation methods in two-sample MR with summary-level data \citep{Burgess2013}. Access to individual-level data would facilitate the use of additional estimation methods \citep{Burgess2017review}, including the popular two-stage least squares algorithm.

In principle, different methods for variable selection and estimation can be combined. For example, the SNPs selected in each iteration of the JAM algorithm could be combined into an allele score, or used for causal effect estimation via the LIML method. Another idea, as hinted in our previous section, would be to combine the variable selection algorithms presented here with pleiotropy-robust MR methods. Although these ideas would be interesting, our goal in this manuscript was to present the current state of the literature, therefore we did not pursue them further.

\section{Simulation Study}

\subsection{Simulation Design}

To compare the various cis-MR approaches, we implemented a simulation study. We generated data under the following simulation model:
\begin{eqnarray} 
	X & = & \sum_{j = 1}^{P} b_{Xj} G_j + \alpha_X U + \epsilon_X \label{model.x} \\
	Y & = & \theta X + \alpha_Y U + \epsilon_Y \label{model.y} \\
	U, \epsilon_X, \epsilon_Y & \sim & N(0, \sigma_0^2) \;\;\;\; \text{independently of each other.} \label{model.s}
\end{eqnarray}
Our simulation design was informed by the real-data applications in the next section. We obtained genetic variants from two gene regions: the \textit{SHBG} region, which encodes the protein sex-hormone binding globulin and is known to be associated with sex hormone traits, and the \textit{HMGCR} region, which encodes the protein HMG coenzyme-A reductase, plays an important role in the metabolic pathway that produces cholesterol and is the drug target of statins. We refer to the real-data applications for more information about these regions. \textit{SHBG} and \textit{HMGCR} represent two distinctly different genetic correlation structures; correlation heat-maps for the two regions are provided in Figures~\ref{hmgcr_plot} and \ref{shbg_plot} below. By basing simulations on two different regions we guarded against spurious observations on comparative performance of the methods that may arise due to subtle properties of the correlation structure in any one region.

We defined the two gene regions as spanning 100Kb pairs on either side of the corresponding gene and used all variants in these regions for which individual-level data were available in the UK Biobank and the minor allele frequencies were higher than $1\%$. This resulted in $P_1 = 717$ genetic variants for the \textit{SHBG} region and $P_2 = 590$ variants for the \textit{HMGCR} region. We extracted individual-level genetic observations for the selected variants from the UK Biobank, using a sample of $N_{ref} = 367643$ non-related individuals of European origin. Aiming to replicate a two-sample Mendelian randomization analysis, we then bootstrapped the UK Biobank data and obtained two genetic matrices $G_1, G_2$ of sizes $N_1, N_2$ respectively. Using the genetic matrix $G_1$, we simulated risk factor measurements according to \eqref{model.x} and used them to obtain $G-X$ summary statistics. Using the matrix $G_2$ and \eqref{model.x}-\eqref{model.y}, we simulated outcome measurements and used them to generate $G-Y$ summary statistics. Finally, the reference correlation matrix was computed using the entire UK Biobank dataset.

To generate risk factor measurements, we used information from our real-data applications and the literature. For the \textit{SHBG} region, we assumed that six independent genetic effects were present in the region. This was inspired by \cite{Coviello2012} who investigated the associations of \textit{SHBG} variants with concentration of the SHBG protein, and suggested that as many as nine independent signals may be present in the region. Three of the nine causal variants identified by the study were not included in our dataset, as they were located more than 100Kb pairs away from the \textit{SHBG} gene, therefore we used the remaining six variants in our simulations. For these six variants, the effects of the risk-increasing allele on the risk factor were generated according to $b_{Xj} \sim |N(0, 0.2)| + 0.1$. This created a pattern of univariate SNP-exposure associations similar to that we observed in our real-data application.

For the \textit{HMGCR} region, we also assumed the existence of six variants independently associated with the risk factor. The six causal signals were placed at genetic variants used in a recent paper \citep{Ference2019} to construct a genetic score for LDL-cholesterol based on the \textit{HMGCR} region. Effects were generated according to $b_{Xj} \sim |N(0, 0.03)| + 0.03$ for the risk-increasing allele; again, this closely resembled the univariate summary statistics obtained in our real-data application. 

We considered three different values for the causal effect parameter: $\theta = 0, 0.05, 0.1$. We set the $G-X$ sample size to be $N_1 = 10000$, which is fairly small for modern MR standards but may be reasonable for a cis-MR study using protein expression data for the risk factor. In supplementary material, we report simulation results using a larger sample size of $N_1 = 100000$. The sample size for SNP-outcome associations was set equal to $N_2 = 180000$, similar to that of the CARDIoGRAMplusC4D dataset \citep{Cardiogram2015} used in our real-data applications. Finally, we assumed that all the genetic variants are valid instruments and did not generate pleiotropic effects.

One of our goals in the simulations was to assess the effect of weak instruments bias on the performance of the various methods. This form of bias can be of particular concern in cis-MR studies, as focusing on a single region means that usually there will be far fewer genetic instruments to use. A common diagnostic for weak instrument bias is to compute the F statistic for the regression of the risk factor on all the genetic variants; a rule-of-thumb is that $F \leq 10$ is indicative of bias.

Therefore, we considered two simulation scenarios. In the first scenario, the proportion of variation in the risk factor explained by the genetic variants in each region was specified according to what has been empirically observed for their protein products, and was set equal to $3\%$ for the \textit{SHBG} region \citep{Jin2012} and $2\%$ for the \textit{HMGCR} region \citep{Krauss2008}. This is a ``strong instruments" setting, in which an ``oracle" MR analysis using only the six causal variants in each region attained an average F statistic of $52$ for the \textit{SHBG} region and $35$ for the \textit{HMGCR} region. In this scenario, the effect of weak instrument bias on the performance of cis-MR methods should be small.

In our second scenario, we reduced the proportion of variation in the risk factor that is explained by the genetic variants to $10\%$ of its value in the previous scenario. For the \textit{SHBG} region, we assumed that the genetic variants only explain $v_G = 0.3\%$ of variation in the risk factor, while for the \textit{HMGCR} region, the genetic variants explained $0.2\%$ of variation. This resulted in a ``weak instruments" scenario in which the ``oracle" MR analysis had an average F statistic of $6.0$ for the \textit{SHBG} region and $4.2$ for the \textit{HMGCR} region.

For each scenario, the simulation was replicated $1000$ times per region and per value of the causal effect.

\subsection{Methods}

In our simulations we implemented the following cis-MR methods:
\begin{itemize}
	\item Single-instrument MR using only the minimum p-value variant in the region.
	\item LD-pruning.
	\item Principal components analysis.
	\item The JAM algorithm.
	\item The F-LIML algorithm. 
	\item The factor-based conditional likelihood ratio (CLR) test.
\end{itemize}
The performance of the conditional method is often quite similar to that of stepwise pruning \citep{Dudbridge2015}, hence the method was not implemented. A brief comparison between conditional analysis, stepwise pruning and PCA was conducted in \cite{Burgess2017}. 

Many of the above methods depend on additional tuning parameters. To assess the sensitivity of each method to the specification of its tuning parameter(s), we used a range of different values. For stepwise pruning, we set a correlation threshold of $\rho = 0.1, 0.3, 0.5, 0.7, 0.9$. For the significance threshold, we use the standard GWAS threshold ($\tau = 5 \times 10^{-8}$) in simulations with ``strong instruments"; with ``weak instruments" there were simulations in which no variants passed the GWAS significance threshold, so we used a lower value ($\tau = 10^{-3}$) instead. For the PCA method, we used principal components that explained $k = 0.99$ or $k = 0.999$ of the variation in the genetic data. For JAM, we implemented a pre-pruning step with a correlation threshold of $\rho = 0.6 ,0.8, 0.9, 0.95$ and no significance threshold - note that JAM does not need a significance threshold for the pre-pruning step because its variable selection discards variants weakly associated with the risk factor anyway. The algorithm was run for 1 million iterations in each instance. Finally for the F-LIML method, we allowed the algorithm to determine the number of latent factors to use.

The simulations were implemented in the statistical software \texttt{R}. Stepwise pruning and correlated-instruments IVW were implemented manually. To implement the PCA approach, we used code provided in the appendix of \cite{Burgess2017}. The JAM algorithm was implemented using the \texttt{R} package \texttt{R2BGLiMS}. The F-LIML algorithm and the summary-data CLR test were implemented using code provided to us by the authors of the relevant publication \citep{Patel2020}. Stepwise pruning was the fastest algorithm to implement and the JAM algorithm was the slowest, although the differences were small on an absolute scale and none of the methods required more than a few seconds to run for each dataset.

The various MR methods in our simulations were subject to two sources of bias. The first source is weak instrument bias, which is known to act towards the direction of the observational association in one-sample MR analyses and towards the null in two-sample MR \citep{Burgess2011, Burgess2016bias}. As discussed earlier, weak instrument bias should be a concern in our second simulation scenario, where the F statistic is below $10$. Note however that even in the first scenario, weak instrument bias may still affect methods using many nuisance variants to obtain a causal effect estimate, as using such variants reduces the value of the F statistic.

Second, the performance of various methods in our simulations can be affected by numerical issues, in particular related to inverting the genetic covariance matrix $G^T G$. Inaccurate estimation of $(G^T G)^{-1}$ may lead to spurious increases or decreases in genetic correlations. This form of bias can be detected by computing the condition number of the genetic covariance matrix, but its direction is not straightforward to assess. It is more likely to affect methods using highly correlated variants, such as stepwise pruning with high correlation thresholds.

\subsection{Results}

Simulation results are reported in Tables \ref{Sim1}-\ref{Sim2}. For each method and each value of the causal effect parameter, we report median causal effect estimates and estimated standard errors. For simulations with $\theta = 0$ we also report the empirical Type I error rates, defined as the proportion of simulations in which the method rejected the null causal hypothesis. For simulations with $\theta \neq 0$ we also report the empirical coverage of confidence intervals and the empirical power to reject the null causal hypothesis. All methods had very high power in the ``strong instruments" scenario with $\theta = 0.1$, and very low power in the ``weak instruments" scenario with $\theta = 0.05$, therefore we do not report power in these two scenarios. Finally, the conditional likelihood ratio test does not provide a causal effect estimate or an associated standard error, hence we only report its coverage, power and Type I error rate.

\begin{table}
\caption{Performance of cis-MR methods in simulations for various values of the causal effect parameter $\theta$, using genetic variants from two gene regions (SHBG and HMGCR) and "strong" genetic instruments (corresponding F statistics $> 10$).}
\label{Sim1}
\begin{footnotesize}
\centering
\begin{tabular}{cc|ccc|cccc|ccc}
  	& 	& \multicolumn{3}{c}{$\theta = 0$} & \multicolumn{4}{c}{$\theta = 0.05$} & \multicolumn{3}{c}{$\theta = 0.1$} \\
  \multicolumn{2}{c}{Method} 	& $\hat{\theta}$ & $se(\hat{\theta})$ & Type I & $\hat{\theta}$ & $se(\hat{\theta})$ & Cov & Power & $\hat{\theta}$ & $se(\hat{\theta})$ & Cov \\ 
  \hline
  \multicolumn{12}{c}{\textit{SHBG} Region} \\
  \hline
    Top SNP & -----	&  0.000 & 0.019 & 0.051 & 0.050 & 0.019 & 0.956 & 0.731 & 0.096 & 0.020 & 0.927 \\ 
  \hline
	Pruning & $\rho = 0.1$ 		& -0.002 & 0.016 & 0.051 & 0.048 & 0.017 & 0.938 & 0.837 & 0.096 & 0.017 & 0.936 \\ 
  	& $\rho = 0.3$ 		& -0.001 & 0.014 & 0.049 & 0.048 & 0.014 & 0.949 & 0.923 & 0.096 & 0.015 & 0.932 \\ 
	& $\rho = 0.5$ 		& -0.001 & 0.013 & 0.051 & 0.047 & 0.014 & 0.943 & 0.941 & 0.095 & 0.014 & 0.927 \\ 
    & $\rho = 0.7$ 		& -0.001 & 0.013 & 0.050 & 0.046 & 0.013 & 0.939 & 0.940 & 0.092 & 0.014 & 0.901 \\ 
  	& $\rho = 0.9$ 		&  0.000 & 0.012 & 0.047 & 0.043 & 0.013 & 0.912 & 0.928 & 0.086 & 0.013 & 0.796 \\ 
  \hline
	PCA & $k = 0.99$ 		& -0.001 & 0.015 & 0.063 & 0.049 & 0.015 & 0.954 & 0.914 & 0.099 & 0.016 & 0.938 \\ 
  	& $k = 0.999$ 		&  0.000 & 0.014 & 0.050 & 0.048 & 0.014 & 0.941 & 0.933 & 0.096 & 0.015 & 0.935 \\ 
  \hline
	JAM & $\rho = 0.6$ 		& -0.001 & 0.014 & 0.048 & 0.048 & 0.014 & 0.954 & 0.935 & 0.095 & 0.015 & 0.938 \\ 
  	& $\rho = 0.8$ 		&  0.000 & 0.014 & 0.049 & 0.047 & 0.014 & 0.957 & 0.934 & 0.095 & 0.015 & 0.939 \\ 
  	& $\rho = 0.9$ 		&  0.000 & 0.014 & 0.041 & 0.048 & 0.014 & 0.955 & 0.932 & 0.094 & 0.015 & 0.935 \\ 
  	& $\rho = 0.95$ 	&  0.000 & 0.014 & 0.045 & 0.047 & 0.014 & 0.956 & 0.933 & 0.094 & 0.015 & 0.933 \\ 
  \hline
  	F-LIML & ----- 		&  0.000 & 0.014 & 0.066 & 0.051 & 0.015 & 0.942 & 0.936 & 0.100 & 0.016 & 0.953 \\ 
  	CLR & -----		&  ----- & ----- & 0.054 & ----- & ----- & 0.947 & 0.926 & ----- & ----- & 0.958 \\ 
  \hline
  \multicolumn{12}{c}{\textit{HMGCR} Region} \\
  \hline
     Top SNP & -----	& 0.001 & 0.018 & 0.048 & 0.049 & 0.019 & 0.953 & 0.776 & 0.099 & 0.019 & 0.924 \\ 
  \hline
	Pruning	& $\rho = 0.1$ 		& 0.001 & 0.018 & 0.048 & 0.049 & 0.019 & 0.953 & 0.776 & 0.099 & 0.019 & 0.924 \\ 
  	& $\rho = 0.3$ 		& 0.001 & 0.017 & 0.048 & 0.048 & 0.018 & 0.948 & 0.795 & 0.097 & 0.018 & 0.928 \\ 
	& $\rho = 0.5$ 		& 0.000 & 0.017 & 0.056 & 0.047 & 0.017 & 0.953 & 0.825 & 0.096 & 0.017 & 0.932 \\ 
    & $\rho = 0.7$ 		& 0.000 & 0.016 & 0.052 & 0.046 & 0.016 & 0.938 & 0.820 & 0.093 & 0.017 & 0.917 \\ 
  	& $\rho = 0.9$ 		& 0.000 & 0.015 & 0.057 & 0.042 & 0.015 & 0.924 & 0.766 & 0.084 & 0.016 & 0.802 \\ 
  \hline
	PCA & $k = 0.99$ 		& 0.000 & 0.017 & 0.049 & 0.049 & 0.017 & 0.949 & 0.830 & 0.099 & 0.018 & 0.930 \\ 
  	& $k = 0.999$ 		& 0.000 & 0.016 & 0.052 & 0.048 & 0.017 & 0.953 & 0.826 & 0.098 & 0.017 & 0.934 \\ 
  \hline
	JAM & $\rho = 0.6$ 		& 0.001 & 0.017 & 0.047 & 0.048 & 0.018 & 0.957 & 0.789 & 0.096 & 0.018 & 0.935 \\ 
  	& $\rho = 0.8$ 		& 0.000 & 0.017 & 0.039 & 0.047 & 0.018 & 0.961 & 0.777 & 0.096 & 0.018 & 0.942 \\ 
  	& $\rho = 0.9$ 		& 0.000 & 0.017 & 0.039 & 0.047 & 0.018 & 0.957 & 0.777 & 0.096 & 0.018 & 0.939 \\ 
  	& $\rho = 0.95$ 	& 0.001 & 0.017 & 0.035 & 0.047 & 0.018 & 0.957 & 0.777 & 0.095 & 0.018 & 0.939 \\ 
  \hline
  	F-LIML & ----- 		& 0.000 & 0.017 & 0.064 & 0.050 & 0.017 & 0.945 & 0.850 & 0.101 & 0.019 & 0.946 \\ 
  	CLR & ----- 		& ----- & ----- & 0.055 & ----- & ----- & 0.947 & 0.834 & ----- & ----- & 0.945 \\ 
  \hline  
\end{tabular}
\end{footnotesize}
\end{table}

Table~\ref{Sim1} contains simulation results from the ``strong instruments" scenario. In this scenario, all methods performed quite well in simulations with a null causal effect. When a positive causal effect was used, PCA, JAM and F-LIML managed to identify the true value of the causal parameter with decent accuracy in both regions. LD-pruning did the same in most cases, but the method's performance deteriorated for large $\rho$ values, exhibiting bias towards the null. The bias was more pronounced for $\theta = 0.1$ than for $\theta = 0.05$. 

The good performance of all methods for $\theta = 0$ suggests that any issues in their performance are due to weak instrument bias, which acts towards the null and hence would only affect simulations with $\theta \neq 0$. Accordingly, any biases observed for $\theta \neq 0$ were towards the null.

For LD-pruning, large values of $\rho$ make the inclusion of weak instruments more likely and the numerical computation and inversion of the correlation matrix more challenging. For small values of $\rho$ the method is more robust to the inclusion of weak instruments. On the other hand, when the genetic region studied contains multiple causal signals, using a small correlation threshold may discard some of the causal variants from the analysis. In our simulations, this translated into a fairly small increase in causal standard errors and a decrease in the method's power to detect a causal effect. We also note that in the simulations of Table~\ref{Sim1}, LD-pruning was implemented after discarding genetic variants whose p-values did not reach genome-wide significance. This is a ``best-case scenario" for the method in terms of avoiding weak instrument bias; in practice, pruning is often implemented using less stringent thresholds \citep{Dudbridge2013}, and the effects of weak instrument bias can be more severe.

As expected, an MR analysis using only the variant with the smallest p-value in the region was unbiased, but had larger standard errors and lower power than other methods. This was more pronounced for the \textit{SHBG} region and less so for the \textit{HMGCR} region, since genetic correlation were stronger in the \textit{HMGCR} region and using only a single variant could partly account for the effects of other variants through correlation. We also note that in our simulations, all causal variants had $G-X$ effects in the same direction and of similar magnitude, and there were no heterogeneous effects towards the outcome. This is a best-case scenario for single-variant analysis; differences between it and other methods are likely to be larger in practice.

The performance of the PCA method was similar for $k = 99\%$ and $k = 99.9\%$ and was quite accurate. With a causal effect of $\theta = 0.1$, the algorithm exhibited a small reduction in coverage due to weak instrument bias but still performed better than LD-pruning. In general, the effect of weak instruments bias on the algorithm is more pronounced for larger values of the tuning parameter $k$. Here, the standard values of $k = 99\%$ and $k = 99.9\%$ worked reasonably well in all simulation scenarios. The power to reject the null causal hypothesis was greater for $k = 99.9\%$, at least for the \textit{SHBG} region.

The JAM algorithm exhibited similar performance to PCA, with a small attenuation of causal effect estimates for $\theta = 0.1$. JAM requires a correlation threshold to be specified for the pruning step before running Bayesian variable selection, but the algorithm's performance was quite robust to the value of that tuning parameter, certainly more so than that of LD-pruning. Its empirical power was slightly higher than PCA for the \textit{SHBG} region, but lower for he \textit{HMGCR} region.

The F-LIML method does not depend on a tuning parameter, as it can automatically determine the number of latent factors to use. Compared to JAM and PCA, the algorithm yielded slightly more accurate causal effect estimates and slightly better calibrated confidence intervals for $\theta = 0.1$ but had slightly inflated Type I error rates for $\theta = 0$. The latter issue was addressed by the conditional likelihood ratio test, at the expense of no causal effect estimates and a slightly lower power than F-LIML.

\begin{table}
\caption{Performance of cis-MR methods in simulations for various values of the causal effect parameter $\theta$, using genetic variants from two gene regions (SHBG and HMGCR) and "weak" genetic instruments (corresponding F statistics $< 10$).}
\label{Sim2}
\begin{footnotesize}
\centering
\begin{tabular}{cc|ccc|ccc|cccc}
  	& 	& \multicolumn{3}{c}{$\theta = 0$} & \multicolumn{4}{c}{$\theta = 0.05$} & \multicolumn{3}{c}{$\theta = 0.1$} \\
  \multicolumn{2}{c}{Method} 	& $\hat{\theta}$ & $se(\hat{\theta})$ & Type I & $\hat{\theta}$ & $se(\hat{\theta})$ & Cov & $\hat{\theta}$ & $se(\hat{\theta})$ & Cov & Power \\ 
  \hline
    \multicolumn{12}{c}{\textit{SHBG} Region} \\
  \hline
     Top SNP & -----	& -0.002 & 0.051 & 0.045 & 0.033 & 0.052 & 0.926 & 0.072 & 0.054 & 0.891 & 0.259 \\      
  \hline
	Pruning & $\rho = 0.1$ 		& -0.001 & 0.047 & 0.045 & 0.031 & 0.048 & 0.922 & 0.069 & 0.050 & 0.868 & 0.290 \\ 
  	& $\rho = 0.3$ 		& -0.002 & 0.040 & 0.045 & 0.032 & 0.041 & 0.921 & 0.066 & 0.042 & 0.843 & 0.359 \\ 
	& $\rho = 0.5$ 		& -0.003 & 0.038 & 0.051 & 0.031 & 0.039 & 0.916 & 0.067 & 0.040 & 0.829 & 0.380 \\ 
    & $\rho = 0.7$ 		& -0.003 & 0.037 & 0.047 & 0.032 & 0.038 & 0.915 & 0.066 & 0.039 & 0.818 & 0.387 \\ 
  	& $\rho = 0.9$ 		& -0.003 & 0.035 & 0.049 & 0.031 & 0.036 & 0.907 & 0.062 & 0.037 & 0.773 & 0.359 \\ 
  \hline
	PCA & $k = 0.99$ 		&  0.000 & 0.040 & 0.050 & 0.035 & 0.041 & 0.924 & 0.068 & 0.042 & 0.876 & 0.392 \\ 
  	& $k = 0.999$ 		& -0.002 & 0.033 & 0.035 & 0.027 & 0.034 & 0.902 & 0.054 & 0.036 & 0.736 & 0.363 \\ 
  \hline
	JAM & $\rho = 0.6$ 		& -0.002 & 0.061 & 0.013 & 0.031 & 0.063 & 0.972 & 0.069 & 0.066 & 0.950 & 0.121 \\ 
  	& $\rho = 0.8$ 		& -0.001 & 0.066 & 0.008 & 0.033 & 0.069 & 0.982 & 0.070 & 0.072 & 0.956 & 0.109 \\ 
  	& $\rho = 0.9$ 		& -0.002 & 0.070 & 0.008 & 0.034 & 0.071 & 0.983 & 0.071 & 0.074 & 0.959 & 0.101 \\ 
  	& $\rho = 0.95$ 	& -0.002 & 0.070 & 0.005 & 0.035 & 0.071 & 0.984 & 0.071 & 0.076 & 0.968 & 0.096 \\ 
  \hline
  	F-LIML & ----- 		& -0.003 & 0.037 & 0.183 & 0.051 & 0.039 & 0.782 & 0.101 & 0.041 & 0.802 & 0.667 \\ 
  	CLR & ----- 		&  ----- & ----- & 0.041 & ----- & ----- & 0.939 & ----- & ----- & 0.948 & 0.404 \\ 
  \hline
    \multicolumn{12}{c}{\textit{HMGCR} Region} \\
  \hline
     Top SNP & -----	& -0.001 & 0.051 & 0.055 & 0.037 & 0.054 & 0.939 & 0.077 & 0.054 & 0.916 & 0.309 \\ 
  \hline
	Pruning & $\rho = 0.1$ 		& -0.001 & 0.050 & 0.055 & 0.037 & 0.053 & 0.939 & 0.077 & 0.054 & 0.914 & 0.308 \\ 
  	& $\rho = 0.3$ 		& -0.002 & 0.047 & 0.049 & 0.039 & 0.050 & 0.942 & 0.072 & 0.051 & 0.893 & 0.309 \\ 
	& $\rho = 0.5$ 		& -0.001 & 0.045 & 0.054 & 0.038 & 0.047 & 0.941 & 0.072 & 0.048 & 0.884 & 0.326 \\ 
    & $\rho = 0.7$ 		& -0.001 & 0.043 & 0.051 & 0.036 & 0.046 & 0.941 & 0.072 & 0.047 & 0.876 & 0.346 \\ 
  	& $\rho = 0.9$ 		&  0.000 & 0.040 & 0.053 & 0.031 & 0.043 & 0.916 & 0.062 & 0.044 & 0.794 & 0.293 \\ 
  \hline
	PCA & $k = 0.99$ 		&  0.003 & 0.047 & 0.054 & 0.040 & 0.049 & 0.946 & 0.080 & 0.051 & 0.919 & 0.359 \\ 
  	& $k = 0.999$ 		&  0.001 & 0.043 & 0.054 & 0.033 & 0.045 & 0.936 & 0.067 & 0.046 & 0.854 & 0.303 \\ 
  \hline
	JAM & $\rho = 0.6$ 		& -0.002 & 0.058 & 0.022 & 0.037 & 0.064 & 0.975 & 0.073 & 0.064 & 0.958 & 0.166 \\ 
  	& $\rho = 0.8$ 		&  0.000 & 0.062 & 0.017 & 0.038 & 0.066 & 0.980 & 0.075 & 0.068 & 0.969 & 0.140 \\ 
  	& $\rho = 0.9$ 		&  0.000 & 0.063 & 0.017 & 0.039 & 0.069 & 0.984 & 0.079 & 0.069 & 0.974 & 0.135 \\ 
  	& $\rho = 0.95$ 	&  0.000 & 0.064 & 0.011 & 0.040 & 0.070 & 0.985 & 0.079 & 0.071 & 0.975 & 0.131 \\ 
  \hline
  	F-LIML & ----- 		&  0.002 & 0.047 & 0.165 & 0.051 & 0.050 & 0.859 & 0.105 & 0.054 & 0.862 & 0.524 \\ 
  	CLR & ----- 		&  ----- & ----- & 0.049 & ----- & ----- & 0.955 & ----- & ----- & 0.952 & 0.331 \\ 
  \hline  
\end{tabular}
\end{footnotesize}
\end{table}

Table~\ref{Sim2} reports simulation results from the ``weak instruments" scenario. Weak instruments bias had a much higher impact in these simulations, with several methods facing attenuation of their causal effect estimates. As expected, any bias occurred only for $\theta \neq 0$ and was towards the null, while all methods also had increased standard errors and low power.

Top-SNP analysis, LD-pruning, PCA and JAM all suffered from weak instrument bias in this scenario. The magnitude of bias was similar for these methods. For pruning and PCA, the bias caused poor coverage properties for confidence intervals and Type I error rates above nominal levels. JAM only selected a small number of genetic variants (it selected an average of $1.6$ variants per run) and attempted to adjust for the presence of weak instruments by producing wider confidence intervals; in fact it was rather conservative in our simulations, with coverage rates above $95\%$, and consequently its power was quite low. In terms of their dependence on tuning parameters, the methods behaved similar to the ``strong instruments" scenario: LD-pruning was more susceptible to bias for large correlation thresholds, while having smaller standard errors. PCA and JAM remained robust to the specification of their tuning parameters.

The F-LIML method had the opposite performance compared to JAM: the algorithm provided quite accurate causal effect estimates, but underestimated standard errors and resulted in confidence intervals with inflated Type I error rates and below nominal coverage. The inefficiency of the algorithm in weak instrument scenarios has been noted by its authors. On the other hand, the conditional likelihood ratio test was the method least affected by weak instruments bias and offered a reliable way of assessing the existence of a causal effect. Its power was worse than the (overprecise) F-LIML method but better than that of the JAM algorithm. The relation between these results and the estimation algorithms used is worth noting. It has been reported in the literature \citep[][Chapter 4]{Angrist2008} that likelihood-based methods can yield near-unbiased causal effect estimates even in the presence of weak instruments, a property not shared by the IVW estimator used by JAM. This was mostly confirmed by our simulation results, although the reduced coverage of the F-LIML approach suggests that even likelihood-based methods are not completely free of weak instrument bias.

In summary, our simulations suggest that simple methods such as LD-pruning and single-variant MR analysis can be reliable in simulations where weak instrument bias is of lesser concern, but should be used with caution in cis-MR analyses where weak instrument bias is suspected. Instrument strength can be assessed using the F statistic, as usual in MR. Even when LD-pruning is used, MR causal effect estimates should be computed and reported for a range of different correlation thresholds as a form of sensitivity analysis. Between JAM, PCA and F-LIML, the differences were small in the ``strong instruments" simulation. With weak instruments, F-LIML provided accurate causal effect estimates but poor uncertainty quantification, JAM yielded biased causal effect estimates, reasonable confidence intervals but low power, the CLR test offered nominal confidence intervals and decent power at the expense of no point estimates and PCA was the method most affected by weak instruments bias. A theoretical advantage of JAM compared to the other approaches is that JAM's variable selection gives an indication of which genetic variants are more likely to be causally associated with the risk factor, although it is not clear whether this additional information would be useful in an MR study where the objective is to perform inference about the $X-Y$ causal effect. Finally, it may be possible to improve the performance of some of the methods by removing very weak variants (e.g. with $p > 0.05$) from the analysis prior to implementing the methods, but we have not considered that here.

\subsection{Additional Simulations}

In supplementary material, we report results from a range of additional simulation scenarios. Specifically, we conducted three additional simulations, on which we comment here briefly.

First, we used a larger sample size of $N_1 = 100000$ from which to obtain risk factor-outcome summary statistics. This could be reminiscent of a cis-MR analysis using a downstream biomarker as a proxy for protein expression, and obtaining $G-X$ summary data from a large-scale GWAS. Our results (Supplementary Table 1) suggested that the various methods perform similarly in this case and issues such as weak instrument bias and robustness to the choice of the pruning threshold are less concerning with large sample sizes.

In the second simulation, we experimented with different numbers of causal variants. We used the \textit{SHBG} region, simulated under the ``strong instruments" scenario, and assumed that the region contains either one or three genetic variants causally associated with the risk factor. The performance of the various methods (Supplementary Table 2) was similar to that observed in our baseline simulations with six causal variants, suggesting that the number of causal signals in the region has little impact on their performance relative to each other, except for the single-variant approach which did expectedly better in simulations with only one causal variant.

In our third supplementary simulation, we assessed the impact of only having access to a small reference dataset on the various methods. In each iteration, we bootstrapped the UK Biobank data in order to obtain a small reference sample of size $N_{ref} = 1000$ and implemented cis-MR using that reference dataset (Supplementary Table 3). The JAM algorithm was the most sensitive method to this change. The algorithm selected more variants on average, exhibited slight differences in its performance for different values of the correlation threshold and performed poorly for $\rho \geq 0.9$. In such datasets, we recommend using a more stringent correlation threshold for the algorithm. The other methods were less affected in comparison.

We refer to the supplement for more details on these simulations.


\section{Applications}

\subsection{Causal Effect of LDL-cholesterol on CHD Risk Using Variants in the \textit{HMGCR} Region}

\subsubsection{Introduction}

We now compare the various cis-MR methods in two real-data applications. In the first application, we use Mendelian randomization to investigate the relationship between low density lipoprotein (LDL) cholesterol concentration and the risk of coronary heart disease (CHD) using genetic variants from the \textit{HMGCR} region. The \textit{HMGCR} region is located in chromosome $5$ and encodes the protein HMG coenzyme-A reductase. The protein plays an important role in the metabolic pathway that produces cholesterol, and its inhibition by statins is a common treatment to reduce LDL-cholesterol levels.

This analysis is performed mainly for illustrative purposes: the causal connection between LDL-cholesterol and CHD risk has already been explored in several papers in the literature. Many of these papers have been genome-wide MR studies \citep{Nitschke2008, Waterworth2010, Burgess2014lipids, Holmes2015lipids, Allara2019}. In the field of cis-MR,  associations between \textit{HMGCR} variants and a range of biomarkers have been used to investigate whether statin treatment increases the risk of type 2 diabetes \citep{Swerdlow2015}. In addition, cis-MR analyses have used the \textit{HMGCR} region as a benchmark in order to assess the suitability of other genetic regions as potential drug targets for CHD risk; this includes the \textit{PCSK9} \citep{Ference2016} and \textit{ACLY} \citep{Ference2019} regions. The \textit{HMGCR} region was also used as an applied example in \cite{Schmidt2019}.

\subsubsection{Datasets and Methods}

Since a clear link between HMGCR protein expression and lowering LDL-cholesterol has been established, we used LDL-cholesterol as a risk factor instead of protein expression. We estimated genetic associations between variants in the \textit{HMGCR} region and LDL-cholesterol using data from the UK Biobank. Genetic associations were computed based on a sample of $N_1 = 349795$ unrelated individuals of European ancestry. Associations with coronary heart disease risk were obtained from the CARDIoGRAMplusC4D consortium \citep{Cardiogram2015}, based on a sample of $N_2 = 184305$ individuals. Finally, genetic correlations were computed using individual-level data for $367643$ unrelated Europeans from the UK Biobank (the sample size for genetic associations with LDL-cholesterol was slightly smaller than the reference sample due to missing values).

\begin{figure}[bt]
\centering
\includegraphics[scale = 0.5]{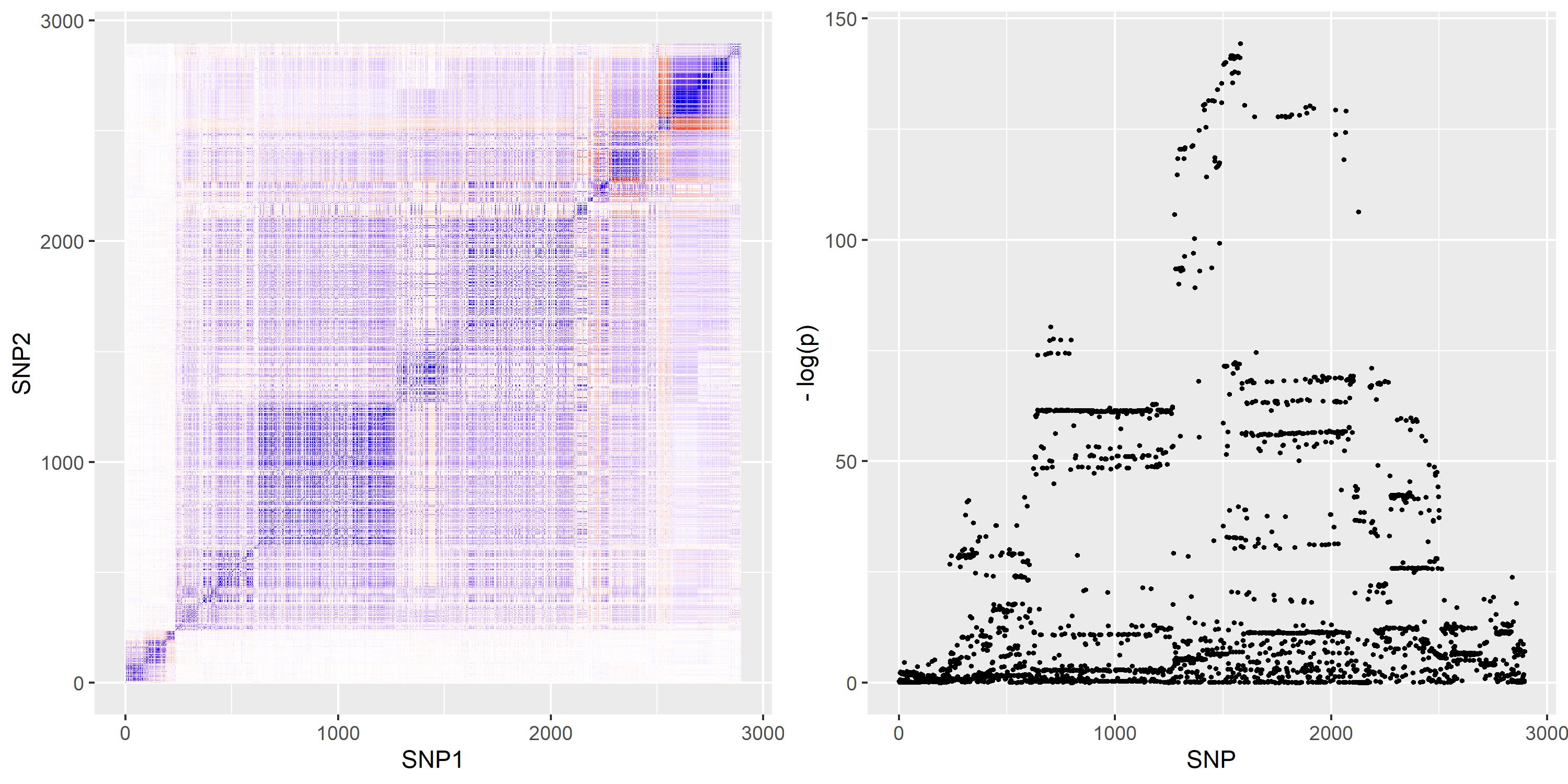}
\caption{Left: genetic correlations in the \textit{HMGCR} region. Right: Manhattan plot of p-values for associations of HMGCR variants with LDL-cholesterol.} \label{hmgcr_plot}
\end{figure}

We used a wider region than in our simulations and extracted information on variants within 500Kb pairs from the \textit{HMGCR} gene. In total, $2915$ variants were present in both the UK Biobank and the CARDIoGRAMplusC4D dataset. Of those $2915$ variants, $20$ were discarded because they either had missing associations with LDL-cholesterol in the UK Biobank dataset or concerned multiple alleles on the same locus that were not detected in both datasets. We did not discard variants with low effect allele frequencies. Our analysis was therefore based on $P = 2895$ genetic variants. In Figure~\ref{hmgcr_plot} we visualize the genetic correlations in the \textit{HMGCR} region, and present a Manhattan plot of genetic association with LDL-cholesterol levels.

We then conducted cis-MR using the minimum p-value variant, LD-pruning, principal components analysis, the JAM algorithm and the factor-based methods. The various methods were implemented using a range of different parameter specifications, similar to those used for our simulation study. For stepwise pruning, we used the values $\rho = 0.1, 0.3, 0.5, 0.7, 0.9$ for the correlation threshold and a GWAS significance threshold of $\tau = 5 \times 10^{-8}$. Of the $2895$ variants in the region, $1424$ had p-values below the GWAS threshold. For the PCA method, we used principal components that explained $k = 99\%$ or $k = 99.9\%$ of variation in the genetic data. For JAM, we implemented a pre-pruning step with a correlation threshold of $\rho = 0.6 ,0.8, 0.9, 0.95$ and no significance threshold. For F-LIML, we allowed the algorithm to determine the number of latent factors to use.

\subsubsection{Results}

\begin{table}[ht]
\caption{Results from various cis-MR methods for the real-data analysis of the effect of LDL-cholesterol on CHD risk, using genetic variants in the \textit{HMGCR} region.}
\label{hmgcr}
\centering
\begin{tabular}{cc|ccc}
  \multicolumn{2}{c}{Method} & $\hat{\theta}$ & $se(\hat{\theta})$ & $95\%$ CI  \\ 
  \hline
  Top-SNP & ----- 		&  0.555 & 0.167 &  (0.228 , 0.882) \\ 
  \hline
  Pruning & $\rho = 0.1$ &  0.389 & 0.136 & (0.122 , 0.656) \\ 
  & $\rho = 0.3$ 		&  0.358 & 0.128 & (0.108 , 0.608) \\ 
  & $\rho = 0.5$ 		&  0.305 & 0.120 & (0.069 , 0.541) \\
  & $\rho = 0.7$ 		&  0.275 & 0.107 & (0.066 , 0.485) \\ 
  & $\rho = 0.9$ 		&  0.378 & 0.069 & (0.242 , 0.513) \\ 
  \hline
  PCA & $k = 0.99$  	& -0.030 & 0.026 & (-0.081 , 0.020) \\ 
  & $k = 0.999$  		&  0.000 & 0.022 & (-0.043 , 0.042) \\ 
  \hline
  JAM & $\rho = 0.6$ 	&  0.401 & 0.168 & (0.072 , 0.731) \\ 
  & $\rho = 0.8$ 		&  0.355 & 0.128 & (0.104 , 0.606) \\ 
  & $\rho = 0.9$ 		&  0.358 & 0.126 & (0.111 , 0.606) \\ 
  & $\rho = 0.95$ 	  	&  0.358 & 0.117 & (0.129 , 0.586) \\ 
  \hline
  F-LIML & -----		&  0.501 & 0.165 &  (0.178 , 0.824) \\ 
  CLR & -----			&  ----- & ----- &  (0.169 , 0.839) \\ 
  \hline
\end{tabular}
\end{table}

Results of our analysis are presented in Table~\ref{hmgcr}. We report causal effect estimates, standard errors and $95\%$ confidence intervals obtained by each method. The reported estimates represent log-odds ratios of increase in CHD risk per standard deviation increase in LDL-cholesterol levels.

Genetically elevated LDL-cholesterol concentration based on variants in the \textit{HMGCR} region is known to be associated with increased risk of coronary heart disease \citep{CTT2010, Ference2019}. This was confirmed by the pruning, JAM, F-LIML and CLR methods. For example, the JAM algorithm using a pre-pruning threshold of $\rho = 0.8$ suggested a log-odds ratio of $0.355$ (odds ratio $1.426$, $95\%$ confidence interval $(1.110, 1.833)$). Interestingly, the principal components method suggested a null effect. The main difference in the implementation of the method compared to our simulation study was the larger number of genetic variants used here. Since the presence of many non-causal variants can exacerbate bias due to weak instruments, we conjecture that this was the issue behind the method's poor performance. To confirm our suspicions, we implemented the PCA method using only the $1424$ GWAS-significant variants in the dataset. The method produced much more reasonable results, in line with other methods: the log-odds ratio estimates were $0.423$ for $k = 99\%$ and $0.448$ for $k = 99.9\%$ and the null causal hypothesis was rejected on both occasions. 

The remaining methods were more consistent in their results. LD-pruning exhibited the usual attenuation towards the null for $\rho = 0.7$ but yielded a larger causal effect estimate for $\rho = 0.9$, as a result of numerical errors. Bias due to numerical errors is more likely to be a concern here than in our simulation study, due to the larger number of variants used. For the JAM algorithm, there were small differences between implementations for $\rho = 0.6$ and larger values of the correlation threshold, but overall the algorithm was more robust to the choice of that threshold than LD-pruning. The algorithm suggested the existence of $3-5$ genetic variants independently associated with LDL-cholesterol in the region. The F-LIML method suggested a slightly higher causal effect than the other methods, although differences were within the margin of statistical error. As an additional form of sensitivity analysis, we implemented both JAM and F-LIML using only GWAS-significant variants; results were similar to those of Table~\ref{hmgcr}.

\subsection{Causal Effect of Testosterone on CHD Risk Using Variants in the \textit{SHBG} Region}

\subsubsection{Introduction}

In our second application, we apply Mendelian randomization using variants in the \textit{SHBG} gene region in order to assess the causal relationship between testosterone levels and coronary heart disease risk. The \textit{SHBG} region is located in chromosome $17$ and encodes sex-hormone binding globulin, a protein that inhibits the function of sex hormones such as testosterone and estradiol. The region has been shown to contain strong genetic associations with testosterone levels, as well as a number of sex hormone traits \citep{Jin2012, Coviello2012, Schooling2018, Ruth2020}. In addition, previous research has suggested that the region is likely to contain several genetic variants independently associated with testosterone, with \cite{Coviello2012} claiming that as many as nine independent signals may be present. This implies that a naive approach using only the variant with the smallest p-value would underestimate the genetic contributions to testosterone levels. The causal relationship between sex hormone traits and CHD risk using variants in the \textit{SHBG} region has been studied by \cite{Burgess2017} and \cite{Schooling2018} with evidence mostly suggesting no causal relationship. Here, we aimed to replicate the analysis of \cite{Burgess2017} using larger datasets.

\subsubsection{Datasets and Methods}

We used serum testosterone levels as the exposure for our MR analysis. We obtained genetic associations with testosterone using summary-level data from the Neale Lab website\footnote{\tiny{\text{https://docs.google.com/spreadsheets/d/1kvPoupSzsSFBNSztMzl04xMoSC3Kcx3CrjVf4yBmESU/edit?ts=5b5f17db\#gid=227859291}}}. These summary data were derived from the UK Biobank, using a sample of $N_1 = 312102$ unrelated individuals of European descent. We defined the genetic region to include genetic variants within 500Kb pairs on either side of the \textit{SHBG} gene. 

\begin{figure}[bt]
\centering
\includegraphics[scale = 0.5]{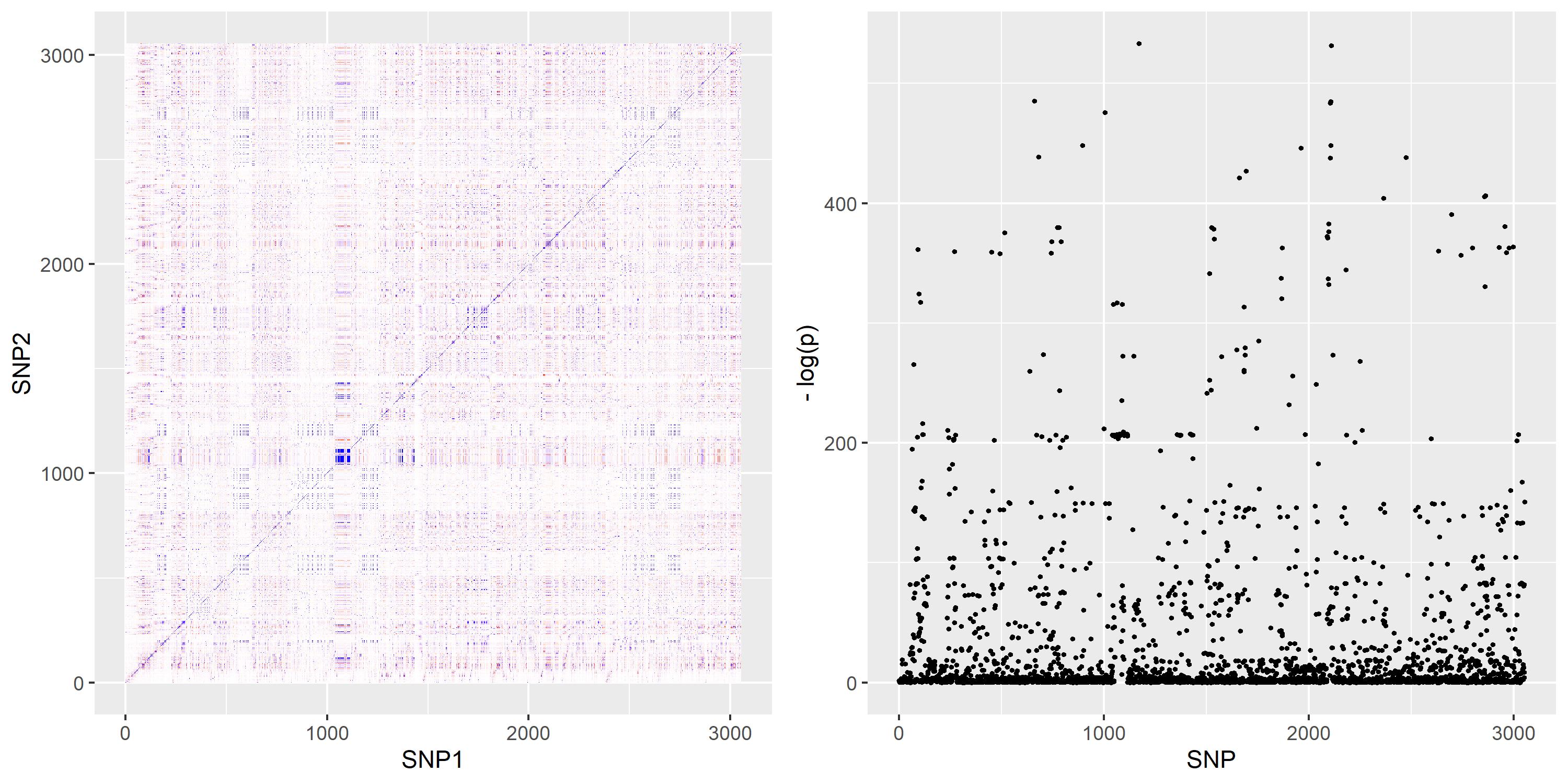}
\caption{Left: genetic correlations in the \textit{SHBG} region. Right: Manhattan plot of p-values for associations of SHBG variants with testosterone levels.} \label{shbg_plot}
\end{figure}

As in the previous application, we obtained genetic associations with CHD risk from the CARDIoGRAMplusC4D dataset \citep{Cardiogram2015}, using a sample of $N_2 = 184305$ individuals. A total of $3053$ genetic variants were present in both datasets and were therefore included in our analysis. We did not conduct separate analyses on males and females, since we did not have access to sex-specific associations with CHD risk. Finally, we used the UK Biobank as a reference dataset from which to extract LD correlations between genetic variants. Figure~\ref{shbg_plot} presents a Manhattan plot of associations with testosterone, as well as the genetic correlations in the region.

We then implemented the various cis-MR methods in order to select genetic instruments and assess whether \textit{SHBG} variants are causally associated with CHD risk. We used the same setting as in the \textit{HMGCR} application for the tuning parameters of each method. The stepwise pruning method was implemented using only variants with GWAS-significant associations with serum testosterone levels. A total of $1156$ variants had p-values lower than the $5 \times 10^{-8}$ threshold. The remaining methods were implemented using all $3053$ variants in the region.

\subsubsection{Results}

\begin{table}[ht]
\caption{Results from various cis-MR methods for the real-data analysis of the effect of serum testosterone levels on CHD risk, using genetic variants in the \textit{SHBG} region.}
\label{shbg}
\centering
\begin{tabular}{cc|ccc}
  \multicolumn{2}{c}{Method} & $\hat{\theta}$ & $se(\hat{\theta})$ & $95\%$ CI  \\ 
  \hline
  Top-SNP & ----- 		& -0.071 & 0.036 & (-0.141 , -0.001) \\ 
  \hline
  Pruning & $\rho = 0.1$ & -0.040 & 0.025 & (-0.088 , 0.008) \\ 
  & $\rho = 0.3$ 		& -0.031 & 0.023 & (-0.076 , 0.014) \\ 
  & $\rho = 0.5$ 		& -0.040 & 0.021 & (-0.081 , 0.001) \\ 
  & $\rho = 0.7$ 		& -0.033 & 0.015 & (-0.062 , -0.003) \\ 
  & $\rho = 0.9$ 		&  0.032 & 0.009 &  (0.014 , 0.051) \\ 
  \hline
  PCA & $k = 0.99$  	& -0.052 & 0.029 & (-0.109 , 0.004) \\ 
  & $k = 0.999$  		& -0.018 & 0.024 & (-0.065 , 0.028) \\ 
  \hline
  JAM & $\rho = 0.6$ 	& -0.042 & 0.024 & (-0.089 , 0.005) \\ 
  & $\rho = 0.8$ 		& -0.037 & 0.024 & (-0.084 , 0.010) \\ 
  & $\rho = 0.9$ 		& -0.040 & 0.024 & (-0.087 , 0.006) \\ 
  & $\rho = 0.95$ 		& -0.035 & 0.024 & (-0.081 , 0.012) \\ 
  \hline
  F-LIML & -----		& -0.037 & 0.025 & (-0.087 , 0.013) \\ 
  CLR & -----			&  ----- & ----- & (-0.087 , 0.014) \\ 
  \hline
\end{tabular}
\end{table}

Table~\ref{shbg} reports causal effect estimates, standard errors and $95\%$ confidence intervals for each method. Once again, estimates are reported in the log-odds ratio scale and represent increases in CHD risk per standard deviation increase in serum testosterone levels. All methods with the exception of top-SNP and LD-pruning at $0.7$ or $0.9$ suggested no causal relationship between testosterone and CHD risk based on the \textit{SHBG} region, although point estimates were consistently in the risk-decreasing direction.

In this example, the performance of the various methods was less seriously affected by weak instruments bias because there was apparently no causal relationship between serum testosterone levels and CHD risk. This is consistent with our simulation design, where no bias was observed under the ``null causal effect" scenario. 

A notable inconsistency was that an MR analysis using only the genetic variant with the smallest p-value in the region suggested a risk-decreasing, statistically significant causal effect. This may be suggestive of heterogeneity in evidence from different variants within the \textit{SHBG} region. Therefore, this example empirically demonstrates the pitfalls of using simplistic single variant analyses when in fact multiple signals exist within a region. The top variant in our analysis was rs1799941, which is known to be associated with testosterone levels \citep{Ruth2020}. 

The LD-pruning method was rather inconclusive in this application. Implementations with a low correlation threshold suggested no causal effect. However, the method suggested an effect in the risk-decreasing direction for $\rho = 0.7$ and in the risk-increasing direction for $\rho = 0.9$. This was combined with a rather sharp increase in precision around the causal estimates. As in our previous application, this is indicative of numerical issues in computing the correlated-instruments IVW estimate and its standard error.

The principal components approach suggested no causal association for both values of its tuning parameter. JAM did the same, and was once again consistent with respect to the value of its tuning parameter. The algorithm suggested the existence of about $8-9$ independent signals in the region, confirming \cite{Coviello2012}. The F-LIML estimate and the CLR confidence interval were in line with the results obtained by JAM.

Our results obtained in this application are similar to those reported by \cite{Burgess2017}. Similar to the results reported in Table~\ref{shbg}, pruning estimates in that paper suggested no causal effect for small correlation thresholds but were unstable for large $\rho$ (in fact, they were more unstable than in our results, possibly due to the smaller sample sizes used in that paper). Estimates from the PCA method were in the risk-decreasing direction but did not achieve statistical significance. Overall, both \cite{Burgess2017} and our analysis suggested no causal association between testosterone and CHD risk based on variants in the \textit{SHBG} region; further evidence of no causality was provided by \cite{Schooling2018}, which also used data from the UK Biobank.

\section{Cis-MR, TWAS and PWAS}

Before concluding our paper, we acknowledge the connections between cis-MR studies and two closely-related approaches that have emerged in the literature in recent years, namely transcriptome-wide association studies (TWAS) and proteome-wide association studies (PWAS). TWAS studies \citep{Gamazon2015, Gusev2016} aim to identify genes associated with an outcome trait of interest, using eQTL variants to instrument gene expression. This is done through a two-step process: first, eQTL variants for a target gene are identified from existing databases and combined to create a genetic score predicting gene expression. And second, the outcome is regressed on predicted expression levels to explore whether the gene relates to the outcome. The original TWAS implementation used individual-level data \citep[PrediXcan, ][]{Gamazon2015}, but methods for summary-level outcome data have been developed since then \citep{Gusev2016, Barbeira2018}.

PWAS studies \citep{Brandes2020} perform effectively the same analysis as TWAS, except they use protein expression data for protein-coding genes instead of gene expression. PWAS use variants that affect the coding regions of genes as instruments and use a similar two-stage process to assess whether protein expression relates to the outcome studied. 

The similarities between cis-MR and TWAS/PWAS studies are obvious. All three approaches aim to identify genetic associations with an outcome of interest at a locus level, and use a similar approach to do so (apart from the difference between using gene expression or protein expression data). In addition, the inference procedure used by TWAS/PWAS methods is effectively a two-stage least squares algorithm. Regarding variable selection, TWAS/PWAS also face the issue of having to select among a group of correlated genetic variants, but these approaches often assume access to individual-level data for gene expression and the corresponding methods are often linked to individual-level eQTL datasets. This gives TWAS methods greater flexibility to overcome the variable selection issues discussed in the current manuscript. 

A few minor differences exist in how TWAS and PWAS studies have been utilized so far in the literature compared to cis-MR studies. For example, cis-MR is typically conducted with a focus on a specific gene region of interest, while TWAS studies are mode agnostic and are often implemented across the genome. In addition, although cis-MR studies regularly use protein expression data as an exposure, the methods developed for cis-MR can be extended to any Mendelian randomization analysis with correlated instruments. However, these differences should not distract from the strong similarities between cis-MR and TWAS/PWAS studies. Although the methods were developed under different names, they are effectively different implementations of the same approach to investigate causal links between gene regions and downstream outcomes.

A detailed description of the methods available for TWAS/PWAS studies is beyond the scope of this manuscript. Instead, we refer to the relevant literature for an overview of methods \citep{Li2021} and challenges \citep{Wainberg2019} for such studies.

\section{Discussion}

Cis-Mendelian randomization is emerging as a widely applicable approach to support drug development and inform clinical trials. In this manuscript we have outlined and compared a range of methods from the cis-MR literature to select variants from a genetic region of interest and estimate the MR causal effect. In particular, we have considered MR using only the minimum p-value variant, and also LD-pruning, which is one of the most commonly used approaches in practice, principal components analysis, factor-based methods and stochastic-search variable selection via the JAM algorithm. We compared the various methods across a range of simulation scenarios, which were based on real genetic correlation structures of two different gene regions drawn from the UK Biobank. We also assessed the performance of the different methods in two cis-MR case studies, investigating the effect of LDL-cholesterol on CHD risk using variants in the \textit{HMGCR} region and the effect of testosterone on CHD risk using variants in the \textit{SHBG} region. 

We have found that LD-pruning can be reliable when using large sample sizes and strong genetic instruments, but can be susceptible to weak instrument bias. Moreover, results obtained using LD-pruning can be sensitive to the correlation threshold used, and using a high correlation threshold can result in numerical instabilities. Based on these results, we would therefore recommend that when the pruning method is used, it is implemented using a range of correlation thresholds as a form of sensitivity analysis. 

Although not completely free from weak instrument bias, methods such as PCA, F-LIML, the JAM algorithm and the conditional likelihood ratio test were less affected by such bias than LD-pruning. JAM was also more robust to the choice of its tuning parameter than pruning, while PCA and F-LIML achieved good performance in a range of different scenarios with the default values for their tuning parameters. Therefore, we recommend that some of these methods are implemented as sensitivity tools in applied cis-MR analyses.

Compared to each other, JAM, PCA, F-LIML and CLR exhibited similar performance in simulations with strong genetic instruments. With weak instruments, the CLR test was least affected by weak instruments bias and provided a valid test for the causal null hypothesis, albeit no point estimates. F-LIML obtained accurate causal effect estimates but yielded confidence intervals with poor coverage, while JAM had biased causal effect estimates but better uncertainty quantification. The principal components IVW approach was affected by weak instruments bias to a greater extent than the other methods, although our real-data applications suggest that PCA can still perform well if very weak variants are removed from the analysis prior to its implementation. 

The use of multiple methods as a form of sensitivity analysis can increase the reliability of results of a cis-MR study. This can be further reinforced by triangulating evidence from different study designs \citep{Gill2020}. In addition to TWAS and PWAS studies discussed earlier, this can include colocalization, which has been shown to yield benefits when used in conjunction with MR to prioritize proteins for drug development \citep{Zheng2019}. Here, we have focused on MR and have not discussed colocalization in detail; we refer to a recent review comparing the two approaches instead \citep{Zuber2022}. 

Our analysis has a number of limitations. Our simulations were by no means exhaustive; for example, they were based on only two genetic regions. In addition, our simulations have not considered other forms of bias that may arise in cis-MR applications. This includes pleiotropic bias, as discussed in Section 2, as well as selection bias, population stratification or winner's curse bias. The latter is worth discussing further, as it could have affected our simulation study. Winner's curse bias occurs when selection of variants into a study is conducted in the same dataset as estimation of instrument-risk factor associations. It has been shown to affect MR studies using independent SNPs as instruments \citep{Sadreev2021}, increasing the magnitude of $G-X$ associations and potentially exacerbating the effects of weak instrument bias. These conclusions also apply to methods that employ variable selection in cis-MR studies. PCA and factor analysis do not perform variable selection explicitly, but may still suffer from a similar type of bias if estimation of the principal component weights or factor loadings is performed in the same dataset as estimation of G-X associations. To address this issue, applied analyses should aim to perform variable selection and estimation of instrument-risk factor associations in separate samples, if possible.

Another limitation is that our work has focused on cis-MR methods that can be implemented using  two-sample summary-level data. Our methods and results can be extended to one-sample summary-data designs, with a few differences (e.g. weak instrument bias acts towards the direction of the observational association in this case) and the additional note that IVW-based methods can suffer from bias due to sample overlap in this case. On the other hand, with access to individual-level data, a range of additional methods such as two-stage least squares and control function approaches can be implemented. These methods might suffer from similar issues as the summary-level methods outlined here, because they also rely on inverting the genetic correlation matrix to perform inference, but may be able to avoid some of the secondary challenges discussed in this paper, such as the potential bias induced by inaccurately estimating the genetic correlation matrix in a reference dataset. It is also possible to compute summary statistics from one-sample individual-level data and then implement the methods presented here, though this may also be biased by sample overlap as mentioned earlier.

Cis-Mendelian randomization is an active area of research, both methodological and applied. As the field develops further, it is likely that new statistical approaches will be created to aid applied investigators in their work. By providing an overview and evaluation of existing methods, we hope that our current paper will contribute to that direction.

\section*{Data Availability Statement}

Access to individual-level UK Biobank data can be obtained upon application to the database at https://www.ukbiobank.ac.uk. For this project, access was granted under Application Number 30931. Summary-level UK Biobank data are also available through the Neale Lab website (http://www.nealelab.is/uk-biobank). Data from the CARDIoGRAMplusC4D consortium are freely available at http://www.cardiogramplusc4d.org/data-downloads.

\section*{Author Contribution Statement}

PJN and SB devised the idea for this project. PJN and AG reviewed the literature and developed the relevant methodology. AG implemented the simulation study. SB facilitated access to real data for the applied analyses. AG conducted the real-data analyses, and all three authors discussed their results. AG led the writing of the manuscript. PJN and SB reviewed the manuscript and provided feedback.

\section*{Conflict of Interest Statement}

Paul Newcombe is currently employed by AstraZeneca. The other authors declare no conflict of interest.

\section*{Ethics Statement}

Both the UK Biobank and the CARDIoGRAMplusC4D consortium have received ethical approval and obtained informed consent from participants prior to data collection. We refer to the websites of these databases for more details.

\section*{Funding and Acknowledgements}

We would like to thank Dr Ashish Patel for many helpful discussions and comments. This work was supported by the UK Medical Research Council (Core Medical Research Council Biostatistics Unit Funding Code: MC UU 00002/7). Apostolos Gkatzionis and Paul Newcombe were supported by a Medical Research Council Methodology Research Panel grant (Grant Number RG88311). Stephen Burgess was supported by a Sir Henry Dale Fellowship jointly funded by the Wellcome Trust and the Royal Society (Grant Number 204623/Z/16/Z). This research has been conducted using the UK Biobank Resource under Application Number 30931.


\bibliography{ReFusion}
\bibliographystyle{chicago}

\pagebreak

\section*{Supplementary Material}

\subsection*{Proportion of Variation Explained by Genetic Instruments}

To fix the proportion $v_G$ of variation in the risk factor explained by the genetic variants, we adjusted the residual variance $\sigma_0^2$ in our simulations accordingly. We first specified a value for $v_G$ and generated SNP-risk factor associations $\beta_{Xj}$ as described in the simulation design section of the manuscript. The value of $\sigma_0^2$ was then set equal to
\begin{equation*}
	\sigma_0^2 = \frac{1 - v_G}{2 v_G} \frac{1}{N_{ref}} \beta_{X}^T G_{ref}^T G_{ref} \beta_{X}
\end{equation*}
where $G_{ref}$ is the reference genetic matrix. This can be justified by recalling that in our simulation design,
\begin{equation*}
	X = \sum_{j = 1}^{P} \beta_{Xj} G_j + \alpha_X U + \epsilon_X
\end{equation*}
with $U, \epsilon_X \sim N(0, \sigma_0^2)$. Taking variances, we have that $\text{Var} (X) = \text{Var} (G \beta_X) + 2 \sigma_0^2$ and therefore,
\begin{eqnarray}
	v_G & = & \frac{\text{Var} (G \beta_X)}{\text{Var} (X)} \, = \, \frac{\text{Var} (G \beta_X)}{\text{Var} (G \beta_X) + 2 \sigma_0^2} \nonumber \\
	& = & \frac{\frac{1}{N_{ref}} \beta_X^T G^T G \beta_X}{\frac{1}{N_{ref}} \beta_X^T G^T G \beta_X + 2 \sigma_0^2} \nonumber
\end{eqnarray}
which yields the previous expression for $\sigma_0^2$. A similar formula was derived in the appendix of \cite{Yang2012}.

\subsection*{Additional Simulations - Large G-X Sample Size}

To augment the analysis presented in the paper, we compared the various cis-MR methods in a range of additional simulations. This section contains simulation results for a simulation using a larger sample size of $N_1 = 100000$ to compute summary-level risk factor-outcome associations. This scenario may represent an applied analysis using a downstream biomarker as a proxy for protein expression, as this would allow researchers to obtain summary-level data from existing large-scale GWAS studies whose sample size often ranges in the hundreds of thousands. This was the case for the two real-data applications presented in the main part of our manuscript.

We conducted simulations both for the SHBG and for the HMGCR region, with six causal variants per region, as in our main simulations. For brevity, we only considered the ``strong instruments" scenario, where $v_G = 3\%$ for the SHBG region and $v_G = 2\%$ for the HMGCR region. Otherwise, the simulations reported here were set up in the same way as those reported in the main part of our paper. Results are reported in Table~\ref{Supp_sim1}.

\begin{table}
\caption{Performance of cis-MR methods in simulations for various values of the causal effect parameter $\theta$, using genetic variants from two gene regions (SHBG and HMGCR), "strong" genetic instruments (corresponding F statistics > 10) and a large $G-X$ sample size ($N_1 = 100000$).}
\label{Supp_sim1}
\begin{footnotesize}
\centering
\begin{tabular}{cc|ccc|cccc|ccc}
  	& 	& \multicolumn{3}{c}{$\theta = 0$} & \multicolumn{4}{c}{$\theta = 0.05$} & \multicolumn{3}{c}{$\theta = 0.1$} \\
  \multicolumn{2}{c}{Method} 	& $\hat{\theta}$ & $se(\hat{\theta})$ & Type I & $\hat{\theta}$ & $se(\hat{\theta})$ & Cov & Power & $\hat{\theta}$ & $se(\hat{\theta})$ & Cov \\ 
  \hline
    \multicolumn{12}{c}{SHBG Region} \\
  \hline
    Top SNP 	& -----	& -0.001 & 0.019 & 0.063 & 0.050 & 0.020 & 0.951 & 0.710 & 0.101 & 0.020 & 0.952 \\ 
  \hline
  	Pruning & $\rho = 0.1$ & -0.001 & 0.016 & 0.054 & 0.050 & 0.016 & 0.943 & 0.868 & 0.100 & 0.016 & 0.954 \\ 
  	& $\rho = 0.3$ 		& -0.001 & 0.014 & 0.039 & 0.050 & 0.014 & 0.946 & 0.940 & 0.099 & 0.015 & 0.958 \\ 
	& $\rho = 0.5$ 		& -0.001 & 0.014 & 0.041 & 0.050 & 0.014 & 0.946 & 0.949 & 0.099 & 0.014 & 0.956 \\ 
    & $\rho = 0.7$ 		& -0.001 & 0.013 & 0.033 & 0.049 & 0.014 & 0.947 & 0.944 & 0.098 & 0.014 & 0.952 \\ 
  	& $\rho = 0.9$ 		& -0.001 & 0.013 & 0.036 & 0.048 & 0.013 & 0.949 & 0.941 & 0.095 & 0.014 & 0.941 \\ 
  \hline
    PCA & $k = 0.99$ 	&  0.000 & 0.015 & 0.038 & 0.051 & 0.015 & 0.950 & 0.909 & 0.100 & 0.016 & 0.957 \\ 
  	& $k = 0.999$ 		&  0.000 & 0.014 & 0.040 & 0.051 & 0.015 & 0.944 & 0.939 & 0.100 & 0.015 & 0.959 \\ 
  \hline
    JAM & $\rho = 0.6$ 	& -0.001 & 0.014 & 0.037 & 0.050 & 0.014 & 0.947 & 0.951 & 0.100 & 0.014 & 0.959 \\ 
  	& $\rho = 0.8$ 		& -0.001 & 0.013 & 0.038 & 0.051 & 0.014 & 0.949 & 0.948 & 0.100 & 0.014 & 0.953 \\ 
  	& $\rho = 0.9$ 		&  0.000 & 0.013 & 0.040 & 0.051 & 0.014 & 0.947 & 0.952 & 0.100 & 0.014 & 0.960 \\ 
  	& $\rho = 0.95$ 	&  0.000 & 0.013 & 0.038 & 0.051 & 0.014 & 0.952 & 0.951 & 0.100 & 0.014 & 0.960 \\ 
  \hline
  	F-LIML & ----- 		& -0.001 & 0.014 & 0.046 & 0.051 & 0.015 & 0.952 & 0.919 & 0.100 & 0.015 & 0.961 \\ 
  	CLR    & ----- 		&  ----- & ----- & 0.044 & ----- & ----- & 0.949 & 0.919 & ----- & ----- & 0.958 \\ 
  \hline
    \multicolumn{12}{c}{HMGCR Region} \\
  \hline
     Top SNP & -----	& 0.000 & 0.019 & 0.055 & 0.050 & 0.019 & 0.950 & 0.738 & 0.100 & 0.020 & 0.951 \\ 
  \hline
    Pruning & $\rho = 0.1$ & 0.000 & 0.018 & 0.048 & 0.050 & 0.018 & 0.956 & 0.769 & 0.100 & 0.019 & 0.956 \\ 
  	& $\rho = 0.3$ 		& 0.000 & 0.017 & 0.051 & 0.049 & 0.018 & 0.947 & 0.795 & 0.099 & 0.018 & 0.951 \\ 
	& $\rho = 0.5$ 		& 0.000 & 0.017 & 0.048 & 0.050 & 0.017 & 0.950 & 0.818 & 0.099 & 0.018 & 0.941 \\ 
    & $\rho = 0.7$ 		& 0.000 & 0.016 & 0.048 & 0.049 & 0.017 & 0.947 & 0.836 & 0.098 & 0.017 & 0.942 \\ 
  	& $\rho = 0.9$ 		& 0.000 & 0.016 & 0.048 & 0.048 & 0.017 & 0.956 & 0.821 & 0.096 & 0.017 & 0.938 \\ 
  \hline
    PCA & $k = 0.99$ 	& 0.000 & 0.017 & 0.047 & 0.050 & 0.017 & 0.956 & 0.815 & 0.100 & 0.018 & 0.947 \\ 
  	& $k = 0.999$ 		& 0.000 & 0.017 & 0.049 & 0.050 & 0.017 & 0.956 & 0.819 & 0.100 & 0.018 & 0.948 \\ 
  \hline
    JAM & $\rho = 0.6$ 	& 0.000 & 0.017 & 0.049 & 0.050 & 0.017 & 0.952 & 0.816 & 0.099 & 0.018 & 0.937 \\ 
  	& $\rho = 0.8$ 		& 0.000 & 0.017 & 0.048 & 0.050 & 0.017 & 0.955 & 0.827 & 0.100 & 0.017 & 0.942 \\ 
  	& $\rho = 0.9$ 		& 0.000 & 0.017 & 0.045 & 0.050 & 0.017 & 0.957 & 0.829 & 0.099 & 0.017 & 0.938 \\ 
  	& $\rho = 0.95$ 	& 0.000 & 0.017 & 0.049 & 0.050 & 0.017 & 0.951 & 0.825 & 0.099 & 0.018 & 0.942 \\ 
  \hline
  	F-LIML & ----- 		& 0.000 & 0.017 & 0.049 & 0.051 & 0.017 & 0.952 & 0.825 & 0.100 & 0.018 & 0.949 \\ 
  	CLR    & ----- 		& ----- & ----- & 0.049 & ----- & ----- & 0.947 & 0.823 & ----- & ----- & 0.947 \\ 
  \hline  
\end{tabular}
\end{footnotesize}
\end{table}

A notable difference in the results compared to our main simulations was that a larger sample size resulted in stronger instruments. A tenfold increase in the sample size resulted in a similar, tenfold increase in the values of the F statistics. For a regression using only the six causal variants, the average F statistic was $514$ for the SHBG and $340$ for the HMGCR region. As a result, the various cis-MR methods were less affected by weak instruments bias and performed quite well. The performance of the stepwise pruning method was was much more consistent for different values of $\rho$ than in the corresponding simulation with $N_1 = 10000$. PCA and JAM were both unaffected by weak instrument bias for $\theta = 0.1$. Confidence intervals constructed using these methods practically attained nominal coverage, and so did the confidence intervals based on F-LIML and CLR. A small reduction in power for single-SNP analysis and LD-pruning with low correlation thresholds were the only real difference between the various methods. This was the case both for the SHBG and for the HMGCR region.

These results imply that weak instrument bias is less likely to affect analyses based on large $G-X$ sample sizes, and stepwise pruning does not underperform other methods in such analyses.

\subsection*{Additional Simulations - Fewer Causal Variants}

In our second set of additional simulations, we modified the number of causal variants that were present in each region. Focusing on the SHBG region, we considered two additional scenarios.

First, we assumed the existence of only a single causal variant in the region, placing the causal signal at the variant that had the smallest univariate p-value in our real-data application (rs1799941). That variant was assumed to have an effect of $\beta_{Xj} = 0.38$ on the risk factor, the same as that observed in our real SHBG-testosterone dataset.

In the second scenario, we generated three independent genetic effects on the risk factor. The effects were placed at genetic variants suggested as independently causal by \cite{Jin2012}, who analysed genetic associations of variants in the SHBG region with serum testosterone levels. Genetic effects for the causal variants were drawn randomly according to $\beta_{Xj} \sim |N(0, 0.2)| + 0.1$ for the risk-increasing allele, same as what we did for the six-causal-variants simulation in the main part of our paper.

Otherwise, the simulations were set up as previously described. We used the ``strong instruments" scenario with $v_G = 3\%$ and a $G-X$ sample size of $N_1 = 10000$. Simulation results are reported in Table~\ref{Supp_sim2}.

\begin{table}
\caption{Performance of cis-MR methods in simulations for various values of the causal effect parameter $\theta$, using genetic variants from the SHBG gene region, "strong" genetic instruments (corresponding F statistics > 10) and either 1 or 3 causal variants.}
\label{Supp_sim2}
\begin{footnotesize}
\centering
\begin{tabular}{cc|ccc|cccc|ccc}
  	& 	& \multicolumn{3}{c}{$\theta = 0$} & \multicolumn{4}{c}{$\theta = 0.05$} & \multicolumn{3}{c}{$\theta = 0.1$} \\
  \multicolumn{2}{c}{Method} 	& $\hat{\theta}$ & $se(\hat{\theta})$ & Type I & $\hat{\theta}$ & $se(\hat{\theta})$ & Cov & Power & $\hat{\theta}$ & $se(\hat{\theta})$ & Cov \\ 
  \hline
    \multicolumn{12}{c}{One Causal Variant} \\
  \hline
     Top SNP & -----	& 0.001 & 0.013 & 0.046 & 0.051 & 0.014 & 0.952 & 0.966 & 0.100 & 0.014 & 0.937 \\      
  \hline
    Pruning & $\rho = 0.1$ & 0.001 & 0.013 & 0.046 & 0.051 & 0.014 & 0.952 & 0.966 & 0.100 & 0.014 & 0.937 \\ 
  	& $\rho = 0.3$ 		& 0.001 & 0.013 & 0.047 & 0.050 & 0.014 & 0.953 & 0.961 & 0.099 & 0.014 & 0.932 \\ 
	& $\rho = 0.5$ 		& 0.001 & 0.013 & 0.045 & 0.049 & 0.014 & 0.949 & 0.956 & 0.096 & 0.014 & 0.920 \\ 
    & $\rho = 0.7$ 		& 0.001 & 0.013 & 0.047 & 0.047 & 0.013 & 0.947 & 0.959 & 0.093 & 0.014 & 0.899 \\ 
  	& $\rho = 0.9$ 		& 0.001 & 0.012 & 0.050 & 0.043 & 0.013 & 0.918 & 0.934 & 0.086 & 0.013 & 0.767 \\ 
  \hline
    PCA & $k = 0.99$ 	& 0.000 & 0.015 & 0.053 & 0.050 & 0.016 & 0.946 & 0.882 & 0.099 & 0.016 & 0.924 \\ 
  	& $k = 0.999$ 		& 0.001 & 0.014 & 0.051 & 0.048 & 0.014 & 0.952 & 0.922 & 0.096 & 0.015 & 0.919 \\ 
  \hline
    JAM & $\rho = 0.6$ 	& 0.001 & 0.013 & 0.046 & 0.051 & 0.014 & 0.952 & 0.965 & 0.100 & 0.014 & 0.938 \\ 
  	& $\rho = 0.8$ 		& 0.001 & 0.013 & 0.046 & 0.051 & 0.014 & 0.952 & 0.966 & 0.100 & 0.014 & 0.938 \\ 
  	& $\rho = 0.9$ 		& 0.001 & 0.013 & 0.046 & 0.051 & 0.014 & 0.953 & 0.966 & 0.100 & 0.014 & 0.938 \\ 
  	& $\rho = 0.95$ 	& 0.001 & 0.013 & 0.046 & 0.051 & 0.014 & 0.953 & 0.966 & 0.100 & 0.014 & 0.938 \\ 
  \hline
  	F-LIML & ----- 		& 0.001 & 0.014 & 0.061 & 0.051 & 0.015 & 0.944 & 0.936 & 0.100 & 0.016 & 0.942 \\ 
  	CLR    & ----- 		& ----- & ----- & 0.047 & ----- & ----- & 0.951 & 0.928 & ----- & ----- & 0.955 \\  
  \hline
    \multicolumn{12}{c}{Three Causal Variants} \\
  \hline
     Top SNP & -----	&  0.000 & 0.016 & 0.051 & 0.050 & 0.016 & 0.956 & 0.852 & 0.098 & 0.017 & 0.934 \\ 
  \hline
    Pruning & $\rho = 0.1$ &  0.000 & 0.015 & 0.047 & 0.049 & 0.016 & 0.953 & 0.868 & 0.097 & 0.016 & 0.928 \\ 
  	& $\rho = 0.3$ 		&  0.000 & 0.014 & 0.046 & 0.049 & 0.014 & 0.942 & 0.919 & 0.096 & 0.015 & 0.930 \\ 
	& $\rho = 0.5$ 		&  0.000 & 0.013 & 0.049 & 0.047 & 0.014 & 0.945 & 0.928 & 0.095 & 0.014 & 0.915 \\ 
    & $\rho = 0.7$ 		&  0.000 & 0.013 & 0.046 & 0.046 & 0.013 & 0.931 & 0.934 & 0.091 & 0.014 & 0.867 \\ 
  	& $\rho = 0.9$ 		&  0.000 & 0.012 & 0.042 & 0.041 & 0.013 & 0.891 & 0.904 & 0.083 & 0.013 & 0.724 \\ 
  \hline
    PCA & $k = 0.99$ 	&  0.000 & 0.015 & 0.051 & 0.049 & 0.015 & 0.941 & 0.884 & 0.096 & 0.016 & 0.932 \\ 
  	& $k = 0.999$ 		&  0.000 & 0.014 & 0.047 & 0.047 & 0.014 & 0.944 & 0.905 & 0.095 & 0.015 & 0.925 \\ 
  \hline
    JAM & $\rho = 0.6$ 	&  0.000 & 0.014 & 0.046 & 0.049 & 0.014 & 0.945 & 0.939 & 0.097 & 0.015 & 0.930 \\ 
  	& $\rho = 0.8$ 		&  0.000 & 0.014 & 0.046 & 0.049 & 0.014 & 0.946 & 0.938 & 0.097 & 0.014 & 0.929 \\ 
  	& $\rho = 0.9$ 		&  0.000 & 0.014 & 0.046 & 0.049 & 0.014 & 0.942 & 0.943 & 0.097 & 0.014 & 0.923 \\ 
  	& $\rho = 0.95$ 	&  0.000 & 0.013 & 0.044 & 0.049 & 0.014 & 0.946 & 0.948 & 0.097 & 0.014 & 0.929 \\ 
  \hline
  	F-LIML & ----- 		&  0.000 & 0.014 & 0.060 & 0.050 & 0.014 & 0.932 & 0.933 & 0.100 & 0.016 & 0.949 \\ 
  	CLR    & ----- 		&  ----- & ----- & 0.049 & ----- & ----- & 0.940 & 0.919 & ----- & ----- & 0.959 \\ 
  \hline
\end{tabular}
\end{footnotesize}
\end{table}

The results closely resembled the ones for our original simulation scenario with ``strong instruments" and six causal variants. All methods were quite accurate for $\theta = 0$. When $\theta \neq 0$, stepwise pruning was subject to weak instrument bias to some extent, especially for large correlation thresholds. The performance of the method depended on the correlation threshold used. PCA and JAM were more consistent in terms of their tuning parameters and were generally quite accurate, while factor-based methods were even more accurate, with a small inflation of Type I error rates under the null for F-LIML.

Most methods performed slightly better with one than with three causal variants, but the differences in their performance were small and it seems that the number of causal signals in the region should have little impact on the choice of which cis-MR method to use. An exception relates to the use of the top-SNP approach, which was expectedly quite accurate when only a single causal variant existed in the region. With three causal variants, top-SNP analysis was subject to the same issues as in our original simulations (namely larger standard errors and lower power than other methods), but to a lesser degree.

\subsection*{Additional Simulations - Small Reference Dataset}

In our third set of additional simulations, we wanted to assess the impact of the reference dataset on the various MR methods. In particular, we assessed whether the use of a small reference dataset, which would imply inaccurate genetic correlation estimates, affects some cis-MR methods worse than others. We therefore repeated the simulations from the main part of our paper, but  bootstrapped the rows of the UK Biobank matrix and only selected $N_{ref} = 1000$ individuals from which to compute genetic correlations, instead of using the entire UK Biobank  dataset of $N_{bb} = 367643$ individuals. In practice, using small datasets as reference data is not uncommon: for example, the 1000 genomes dataset is commonly used for this purpose. The use of a small reference dataset does not systematically bias the estimated genetic covariance matrix, as would be the case with population stratification.

We used both genetic regions and the ``strong instruments" scenario. Again, the simulation design was identical to those reported in the main part of the paper, except using a smaller reference dataset. Results are reported in Table~\ref{Supp_sim3}.

\begin{table}
\caption{Performance of cis-MR methods in simulations for various values of the causal effect parameter $\theta$, using genetic variants from two gene regions (SHBG and HMGCR), "strong" genetic instruments (corresponding F statistics > 10) and a smaller reference sample ($N_{ref} = 1000$).}
\label{Supp_sim3}
\begin{footnotesize}
\centering
\begin{tabular}{cc|ccc|cccc|ccc}
  	& 	& \multicolumn{3}{c}{$\theta = 0$} & \multicolumn{4}{c}{$\theta = 0.05$} & \multicolumn{3}{c}{$\theta = 0.1$} \\
  \multicolumn{2}{c}{Method} 	& $\hat{\theta}$ & $se(\hat{\theta})$ & Type I & $\hat{\theta}$ & $se(\hat{\theta})$ & Cov & Power & $\hat{\theta}$ & $se(\hat{\theta})$ & Cov \\ 
  \hline
    \multicolumn{12}{c}{SHBG Region} \\
  \hline
     Top SNP & -----	& -0.001 & 0.019 & 0.039 & 0.048 & 0.019 & 0.941 & 0.676 & 0.095 & 0.020 & 0.917 \\ 
  \hline
    Pruning & $\rho = 0.1$ & -0.001 & 0.016 & 0.046 & 0.048 & 0.016 & 0.925 & 0.799 & 0.095 & 0.017 & 0.892 \\ 
  	& $\rho = 0.3$ 		&  0.000 & 0.014 & 0.045 & 0.048 & 0.014 & 0.942 & 0.900 & 0.095 & 0.015 & 0.901 \\ 
	& $\rho = 0.5$ 		&  0.000 & 0.013 & 0.049 & 0.047 & 0.014 & 0.933 & 0.914 & 0.093 & 0.014 & 0.892 \\ 
    & $\rho = 0.7$ 		&  0.000 & 0.013 & 0.048 & 0.046 & 0.013 & 0.917 & 0.916 & 0.090 & 0.014 & 0.863 \\ 
  	& $\rho = 0.9$ 		&  0.000 & 0.012 & 0.057 & 0.043 & 0.012 & 0.871 & 0.900 & 0.084 & 0.013 & 0.750 \\ 
  \hline
    PCA & $k = 0.99$ 	&  0.000 & 0.015 & 0.041 & 0.049 & 0.015 & 0.939 & 0.883 & 0.097 & 0.015 & 0.922 \\ 
  	& $k = 0.999$ 		&  0.000 & 0.014 & 0.052 & 0.048 & 0.014 & 0.925 & 0.904 & 0.094 & 0.015 & 0.919 \\ 
  \hline
    JAM & $\rho = 0.6$ 	&  0.000 & 0.014 & 0.038 & 0.047 & 0.014 & 0.937 & 0.896 & 0.093 & 0.015 & 0.906 \\ 
  	& $\rho = 0.8$ 		&  0.000 & 0.014 & 0.039 & 0.047 & 0.014 & 0.922 & 0.896 & 0.092 & 0.014 & 0.887 \\ 
  	& $\rho = 0.9$ 		&  0.000 & 0.014 & 0.063 & 0.047 & 0.014 & 0.910 & 0.895 & 0.092 & 0.014 & 0.867 \\ 
  	& $\rho = 0.95$ 	&  0.000 & 0.014 & 0.093 & 0.047 & 0.014 & 0.888 & 0.885 & 0.091 & 0.014 & 0.843 \\ 
  \hline
  	F-LIML & ----- 		&  0.000 & 0.014 & 0.052 & 0.050 & 0.014 & 0.921 & 0.907 & 0.099 & 0.016 & 0.929 \\ 
  	CLR    & ----- 		&  ----- & ----- & 0.040 & ----- & ----- & 0.929 & 0.895 & ----- & ----- & 0.933 \\ 
  \hline
    \multicolumn{12}{c}{HMGCR Region} \\
  \hline
     Top SNP & -----	& -0.001 & 0.018 & 0.047 & 0.048 & 0.018 & 0.951 & 0.753 & 0.099 & 0.019 & 0.921 \\ 
  \hline
    Pruning & $\rho = 0.1$ & -0.001 & 0.018 & 0.048 & 0.048 & 0.018 & 0.951 & 0.757 & 0.099 & 0.019 & 0.919 \\ 
  	& $\rho = 0.3$ 		& -0.001 & 0.017 & 0.056 & 0.047 & 0.017 & 0.951 & 0.770 & 0.097 & 0.018 & 0.928 \\ 
	& $\rho = 0.5$ 		& -0.001 & 0.016 & 0.052 & 0.047 & 0.017 & 0.947 & 0.807 & 0.096 & 0.017 & 0.920 \\ 
    & $\rho = 0.7$ 		& -0.001 & 0.016 & 0.052 & 0.046 & 0.016 & 0.943 & 0.818 & 0.093 & 0.017 & 0.893 \\ 
  	& $\rho = 0.9$ 		&  0.000 & 0.014 & 0.113 & 0.042 & 0.015 & 0.836 & 0.785 & 0.084 & 0.015 & 0.735 \\ 
  \hline
    PCA & $k = 0.99$ 	& -0.001 & 0.017 & 0.047 & 0.048 & 0.017 & 0.950 & 0.816 & 0.098 & 0.018 & 0.921 \\ 
  	& $k = 0.999$ 		&  0.000 & 0.016 & 0.051 & 0.047 & 0.017 & 0.949 & 0.814 & 0.097 & 0.017 & 0.922 \\ 
  \hline
    JAM & $\rho = 0.6$ 	& -0.001 & 0.017 & 0.047 & 0.047 & 0.018 & 0.949 & 0.782 & 0.096 & 0.018 & 0.933 \\ 
  	& $\rho = 0.8$ 		& -0.001 & 0.017 & 0.062 & 0.046 & 0.017 & 0.934 & 0.794 & 0.094 & 0.018 & 0.910 \\ 
  	& $\rho = 0.9$ 		&  0.000 & 0.017 & 0.093 & 0.045 & 0.017 & 0.895 & 0.790 & 0.093 & 0.018 & 0.859 \\ 
  	& $\rho = 0.95$ 	& -0.001 & 0.017 & 0.150 & 0.044 & 0.017 & 0.824 & 0.759 & 0.092 & 0.018 & 0.791 \\ 
  \hline
  	F-LIML & ----- 		& -0.001 & 0.016 & 0.052 & 0.049 & 0.017 & 0.941 & 0.825 & 0.101 & 0.019 & 0.945 \\ 
  	CLR    & -----		&  ----- & ----- & 0.049 & ----- & ----- & 0.945 & 0.810 & ----- & ----- & 0.942 \\ 
  \hline
\end{tabular}
\end{footnotesize}
\end{table}

The JAM algorithm proved to be more sensitive to the reference dataset. The algorithm adjusts marginal SNP-trait associations using the reference genetic correlations. By using an inaccurate correlation pattern, the algorithm's adjustment was adversely affected and this resulted in selecting more genetic variants than in runs with a larger reference dataset. This was especially the case for large values of the correlation threshold, in which case the pre-pruning step only discards a small number of variants before running JAM. For example, for the SHBG region with $\theta = 0$ and $\rho = 0.95$, the algorithm had a posterior model size of $9.27$ compared to $3.75$ when the entire UK Biobank was used as a reference dataset. This meant that the algorithm was more susceptible to weak instrument bias, made JAM causal effect estimates slightly more variable for larger $\rho$ values and increased Type I error rates. For smaller values of the correlation threshold the algorithm's performance was affected less.

The performance of LD-pruning also attenuated for large values of its correlation threshold, but this attenuation was in line with what we observed in our original simulations, using a large reference dataset. Likewise, principal components analysis and factor-based methods seemed fairly robust to using a smaller reference dataset.

\end{document}